\documentclass{JHEP3}
\usepackage{epsfig}
\usepackage{graphicx}
\usepackage{subfigure,caption}
\newcommand{\D}{\Delta}
\renewcommand{\d}{\delta}
\renewcommand{\l}{\lambda}
\renewcommand{\L}{\Lambda}
\renewcommand{\b}{\beta}
\renewcommand{\a}{\alpha}
\newcommand{\p}{\phi}
\renewcommand{\k}{\kappa}

\newcommand{\g}{\gamma}

\newcommand{\m}{\mu}
\renewcommand{\r}{\rho}

\newcommand{\s}{\sigma}

\newcommand{\CD}{{\cal D}}

\newcommand{\E}{{\cal E}}

\newcommand{\tU}{\widetilde{U}}

\newcommand{\J}{{\cal J}}

\newcommand{\e}{\epsilon}

\newcommand{\oh}{{\textstyle{\frac{1}{2}}}}

\newcommand{\dg}{\dagger}
\newcommand{\non}{\nonumber}

\newcommand{\rf}[1]{(\ref{#1})}
\newcommand{\ra}{\rightarrow}
\newcommand{\pa}{\partial}

\title{Center vortices and the Gribov horizon}
\author{Jeff Greensite \\
The Niels Bohr Institute, Blegdamsvej 17, DK-2100 Copenhagen \O,
Denmark,\\
Physics and Astronomy Dept., San Francisco State University, San
Francisco, CA~94117, USA.\\
E-mail: \email{greensit@stars.sfsu.edu}}
\author{{\v S}tefan Olejn\'{\i}k \\
Institute of Physics, Slovak Academy of Sciences, SK--845 11
Bratislava, Slovakia.\\
E-mail: \email{stefan.olejnik@savba.sk}}
\author{Daniel Zwanziger \\
Physics Department, New York University, New York, NY~10003, USA.\\
E-mail: \email{daniel.zwanziger@nyu.edu}}
\date{\today}
\abstract{
    We show how the infinite color-Coulomb energy
of color-charged states is related to enhanced density of near-zero
modes of the Faddeev--Popov operator,
and calculate this density numerically for both pure Yang--Mills and
gauge-Higgs systems at zero temperature, and for pure gauge theory
in the deconfined phase. We find that the enhancement of the
eigenvalue density is tied to the presence of percolating center
vortex configurations, and that this property disappears when center
vortices are either removed from the lattice configurations, or
cease to percolate.  We further demonstrate that thin center
vortices have a special geometrical status in gauge-field
configuration space: Thin vortices are located at conical or wedge
singularities on the Gribov horizon.  We show that the Gribov region
is itself a convex manifold in lattice configuration space.  The
Coulomb gauge condition also has a special status; it is shown to be
an attractive fixed point of a more general gauge condition,
interpolating between the Coulomb and Landau gauges.}
\keywords{Confinement, Lattice Gauge Field Theories, Solitons
Monopoles and Instantons}
\begin{document}
\maketitle
%
% Section I
%
\section{Introduction}\label{Introduction}

     Of the many different ideas that have been advanced to explain quark
confinement, more than one may be right, or at least partially
right, and these should be related in some way.
As in the old story of six blind men describing an elephant,
those of us concerned with the QCD confinement mechanism might benefit from
unifying some of our separate impressions, to form a better
image of the entire beast.

   In this article we investigate the relationship between center vortices
(for a review, cf. ref.\ \cite{review}) and the Gribov horizon in
Coulomb gauge, whose relevance to confinement has been advocated
by Gribov and Zwanziger \cite{horizon}.  We begin with the simple
fact that in a confining theory, the energy of an isolated color charge
is (infrared) infinite, and this energy is a lower bound on the
color-Coulomb energy \cite{Dan} (defined below).  This fact implies an
enhancement of the density of near-zero eigenvalues $\l_n$
\begin{equation}
        M \p^{(n)} = \l_n \p^{(n)}
\label{eigen}
\end{equation}
of the Faddeev--Popov (F-P) operator in Coulomb gauge
\begin{equation}
        M = - \nabla \cdot \CD(A),
\end{equation} where $\CD(A)$ is the covariant derivative. The F-P
eigenvalue density is an observable which we are able to calculate
numerically (section \ref{fped}), via lattice Monte Carlo and
sparse-matrix techniques. Applying standard methods \cite{dFE} we
are able to separate any thermalized lattice configuration into
vortex-only and vortex-removed components, and we find that the
enhancement of the F-P eigenvalue density can be entirely attributed
to the vortex component of the gauge fields.  Vortices are
associated with an enhancement in the density of F-P eigenvalues
$\l$ near $\l=0$; this enhancement is key to the divergence of the
color-Coulomb energy in Coulomb gauge. It is absent in the Higgs
phase of a gauge-Higgs system (section \ref{ghiggs}), where remnant
gauge symmetry is broken \cite{Us}, and vortices cease to percolate.
In particular, we compare the F-P eigenvalue density found in the
Higgs phase to the corresponding density for configurations in the
confined phase, with vortices removed. These densities are identical
in form. The density of F-P eigenvalues in the high-temperature
deconfined phase of pure gauge theory is examined in section
\ref{hight}.  We find that the linearly rising, unscreened
color-Coulomb potential, which is present in the deconfined phase
\cite{Us}, is associated with an enhanced density of F-P
eigenvalues, and that this enhancement is again attributable to the
vortex component of the gauge field.  Divergence of the
color-Coulomb energy is a necessary but not sufficient condition for
color confinement, and this phase provides an interesting example
where the infinite color-Coulomb potential gets screened.  Although
the horizon scenario was invented to describe confinement, it nicely
accounts for the divergence of the color-Coulomb energy in the
deconfined phase, as explained in the Conclusion. It is also shown
(section \ref{tvort}) how an array of center vortices, set up ``by
hand" to simulate some aspects of a percolating vortex
configuration, leads to an accumulation of F-P eigenvalues near
$\l=0$.

   In section \ref{vv} we demonstrate that center configurations
(equivalently: thin center vortices) have some remarkable geometrical
properties in lattice configuration space.  First, as already shown in
ref.\ \cite{Us}, center configurations lie on the Gribov horizon.  It
is known that the Gribov horizon is a convex manifold for continuum
gauge fields \cite{semenov} $-$ a result which we extend here (section
\ref{convex}) to lattice gauge fields $-$ and one might therefore
suspect that the Gribov horizon is also smooth and differentiable.  In
fact, thin vortex configurations turn out to be distinguished points
on the Gribov horizon, where the manifold acquires conical or ``wedge"
singularities.  Finally, in section \ref{cgfp}, we point out that the
Coulomb gauge condition also has a special status, in that Coulomb
gauge is an attractive fixed point of a more general interpolating
gauge condition, which has the Coulomb and Landau gauge conditions as
special cases.

\section{The F-P eigenvalue density}\label{fped}

   The energy of an isolated color charge is
(infrared) infinite in the confined phase, even on the lattice
where ultraviolet divergences are regulated.  We will consider charged
states in Coulomb gauge, which, for a single static point charge,
has the simple form
\begin{equation}
        \Psi^\a_C[A;x] = \psi^\a(x) \Psi_0[A],
\label{charge1}
\end{equation}
where $\a$ is the color index for a point charge in color group
representation $r$, and $\Psi_0$ is the Coulomb gauge ground
state.  The gauge-field excitation energy $\E_r$ of this state,
above the ground state energy, is due entirely to the
non-local part of the hamiltonian
\begin{equation}
\label{excitation}
        \E_r = {\langle \Psi^\a_C[A;x] | H_{coul} | \Psi^\a_C[A;x] \rangle
       \over  \langle \Psi^\a_C[A;x] | \Psi^\a_C[A;x] \rangle}
             - \langle \Psi_0 | H_{coul} | \Psi_0 \rangle.
\end{equation}
We recall that in Coulomb gauge, the hamiltonian is a sum
$H=H_{glue}+H_{coul}$, where
\begin{eqnarray}
        H_{glue} = \oh \int d^3x\;( \J^{-\oh}\vec{E}^{{\rm tr},a} {\cal J}
        \cdot \vec{E}^{{\rm tr},a} \J^{-\oh} + \vec{B}^a \cdot \vec{B}^a),
\non \\
        H_{coul} = \oh \int d^3x d^3y\;\J^{-\oh}\r^a(x) \J
                K^{ab}(x,y;A) \r^b(y) \J^{-\oh},
\non
\end{eqnarray}
and
\begin{eqnarray}
        K^{ab}(x,y;A) &=& \left[ M^{-1}
        (-\nabla^2) M^{-1} \right]^{ab}_{xy},
\non \\
        \r^a &=& \r_q^a - g f^{abc} A^b_k E^{{\rm tr},c}_k,
\non \\
         \J &=& \det[-\nabla \cdot \CD(A)].
\end{eqnarray}
Then the excitation energy of the charged state, in SU($N$) gauge theory, is
\begin{equation}
          \E_r = g^2 {1\over d_r} \mbox{Tr}[L^a L^b]
                      \langle K^{ab}(x,x;A) \rangle
             = g^2 C_r {1\over N^2 -1} \langle K^{aa}(x,x;A) \rangle,
\end{equation}
where the $\{L^a\}$ are color group generators, $d_r$ the dimension,
and $C_r$ the quadratic Casimir of representation $r$ of the color
charge.  Therefore, the excitation energy $\E_r$ is the energy of the
longitudinal color electric field due to the static source, which we
can identify as the color Coulomb self-energy.  This energy is
ultraviolet divergent in the continuum, but of course that divergence
can be regulated with a lattice cut-off.  The more interesting point is
that in a confining theory, $\E_r$ must still be divergent at infinite
volume, even after lattice regularization, due to infrared effects.

The excitation energy \rf{excitation} represents the
(infrared-infinite) energy of {\it unscreened} color charge in the
state \rf{charge1}, which in general is a highly excited state that is
not an eigenstate of the hamiltonian.  States of this kind are useful
for extracting the self-energy of an isolated charge due to its
associated color-Coulomb field.  On the other hand, the \emph{minimal}
free energy $\E_s$ of a state containing a static external charge, at
inverse temperature $T$, is obtained from the value of the Polyakov
loop $P \sim \exp(- \E_s T)$.  This minimal energy may be infrared
finite in an unconfined phase even if $\E_r$ is not, providing the
external charge can be screened by dynamical matter fields, or by
high-temperature effects.  The infrared divergence of $\E_r$ must be
understood as a necessary, but not sufficient, condition for
confinement.

    We note in passing that the charged state \rf{charge1} in Coulomb
gauge corresponds, in QED in temporal gauge, to the well-known form
\begin{equation}
        \Psi^{QED}[A;x] = \exp\left[ie\int d^3z ~ A(z) \cdot \nabla
                        {1\over 4 \pi|x-z|} \right] \psi(x)
                        \Psi_0^{QED}[A]
\label{QEDcharge}
\end{equation}
The investigation of this type of ``stringless" state with external
charges in non-abelian theories, using perturbative methods, was
undertaken some time ago by Lavelle and McMullan in ref.\
\cite{lavelle}.  The exponential prefactor in eq.\ \rf{QEDcharge} can
be identified as the gauge transformation taking an arbitrary
configuration $A_k(x)$ into Coulomb gauge.  This feature generalizes
to non-abelian theories, and ``stringless" states with static charges
in temporal gauge can be formally expressed in terms of the gauge
transformation to Coulomb gauge, as shown in ref.\ \cite{Us}.

     We now proceed to the lattice formulation, with an SU(2) gauge group.
The link variables can be expressed as
\begin{equation}
        U_\m(x) = b_\m(x) + i \vec{\s}\cdot \vec{a}_\m(x) ~,~~~ b_\m(x)^2 +
                                                     \sum_c a^c_\m(x)^2 = 1
\end{equation}
and (when the lattice version of the Coulomb gauge condition
$\nabla \cdot A = 0$ is satisfied)
  the lattice Faddeev--Popov operator is
\begin{eqnarray}
  M^{ab}_{xy} &=& \d^{ab} \sum_{k=1}^3\left\{ \d_{xy}  \left[b_k(x)
    + b_k(x-\hat{k})\right]
    - \d_{x,y-\hat{k}} b_k(x)
    - \d_{y,x-\hat{k}} b_k(y) \right\}
\non \\
       &-& \epsilon^{abc} \sum_{k=1}^3\left\{ \d_{x,y-\hat{k}} a^c_k(x)
                  - \d_{y,x-\hat{k}} a^c_k(y)  \right\},
\label{M}
\end{eqnarray}
where indices $x,y$ denote lattice sites at fixed time.  Denote the
Green's function corresponding to the F-P operator as\footnote{This
expression assumes $M$ is invertible, a point which will be
discussed in subsection \ref{observables}, below.}
\begin{equation}
          G^{ab}_{xy} = \left[M^{-1}\right]^{ab}_{xy}
                      = \sum_n {\phi^{a(n)}_x \phi^{b(n)*}_y \over \l_n},
\end{equation}
where $\phi^{a(n)}_x,~\l_n$ are the $n$-th normalized eigenstate and
eigenvalue of the lattice F-P operator $M_{xy}^{ab}$.  Defining the
representation-independent factor $\E\equiv \E_r/(g^2 C_r)$ in the Coulomb
self-energy, we find
\begin{eqnarray}
      \E &=&  {1\over N^2 -1} \langle K^{aa}(x,x;U) \rangle
\non \\
         &=&  {1\over 3} \sum_{y_1 y_2} \langle
        G_{xy_1}^{ab}(-\nabla^2)_{y_1 y_2} G_{y_2x}^{ba} \rangle
         =  {1\over 3V_3}  \sum_x \sum_{y_1 y_2} \langle
              G_{xy_1}^{ab}(-\nabla^2)_{y_1 y_2}
                      G_{y_2x}^{ba} \rangle
\non \\
         &=&  {1\over 3V_3}   \sum_x \sum_{y_1 y_2} \sum_m \sum_n \left\langle
          {\phi^{a(m)}_x \phi^{b(m)*}_{y_1} \over \l_m} (-\nabla^2)_{y_1 y_2}
          {\phi^{b(n)}_{y_2} \phi^{a(n)*}_x \over \l_n} \right\rangle
\non \\
         &=&   {1\over 3V_3}   \sum_{y_1 y_2} \sum_n \left\langle
          {\phi^{a(n)*}_{y_1} (-\nabla^2)_{y_1 y_2}
           \phi^{a(n)}_{y_2} \over \l_n^2} \right\rangle,
\end{eqnarray}
where $V_3=L^3$ is the lattice 3-volume. Also defining
\begin{equation}
          F_n = \sum_{y_1 y_2} \phi^{a(n)*}_{y_1} (-\nabla^2)_{y_1 y_2}
           \phi^{a(n)}_{y_2}
              = \vec{\p}^{(n)*} \cdot (-\nabla^2) \vec{\p}^{(n)}
\end{equation}
we have
\begin{equation}
       \E = {1\over 3V_3} \sum _n \left\langle
               {F_n \over \l_n^2} \right\rangle.
\end{equation}

    The Faddeev--Popov operator, on the lattice, is a $3V_3 \times 3V_3$ sparse
matrix; the number of linearly-independent eigenstates is therefore $3V_3$.
Let $N(\l,\l+\D\l)$ be the number of eigenvalues in the range
$[\l,\l+\D\l]$.  We define,
on a large lattice, the normalized density of eigenvalues
\begin{equation}
      \r(\l) \equiv {N(\l,\l+\D\l) \over 3V_3 \D \l}.
\label{density}
\end{equation}
Then as the lattice volume tends to infinity,
\begin{equation}
      \E = \int_0^{\l_{max}} {d\l  \over \l^2} ~ \r(\l) F(\l) ,
\end{equation}
where it is understood that the integrand is averaged over the
ensemble of configurations.
From this we derive a condition for the confinement phase:
The excitation energy $\E_r = g^2 C_r \E$ of a static, unscreened
color-charge is divergent if, at infinite volume,
\begin{equation}
           \lim_{\l \ra 0} { \r(\l) F(\l) \over \l} > 0.
\label{condition}
\end{equation}
In perturbation theory, at zero-th order in the gauge-coupling, the
Faddeev--Popov operator is simply a laplacian, whose eigenstates are
plane waves.  Then $\l=k^2$, where $\vec{k}$ is the momentum, and from
the volume element of momentum space it is easy to see that to zeroth
order, in the limit of infinite lattice volume,
\begin{equation}
         \r(\l) = {\l^{1/2} \over 4\pi^2} ~,~~ F(\l) = \l ,
\label{zeroth}
\end{equation}
which obviously does not satisfy the confinement condition
\rf{condition}.  At zeroth-order
we have $\E = \l_{max}^{1/2}/(2\pi^2)$.

    We now recall briefly some aspects of the Gribov horizon
scenario \cite{horizon}.  The lattice version of the Coulomb
gauge-fixing condition $\nabla \cdot A = 0$ is satisfied by any
lattice configuration $U$ such that
\begin{equation}
        R = \sum_x \sum_{k=1}^3 \mbox{Tr}[g(x)U_k(x)g^{-1}(x+\widehat{k})]
\end{equation}
is stationary at the trivial gauge transformation $g=I$.  The
Faddeev--Popov operator is obtained from the second derivative of
$R$ with respect to gauge transformations, so if $R$ is a local
maximum, then all the eigenvalues of the Faddeev--Popov operator are
positive.  The subspace of configuration space satisfying this
condition is known as the Gribov region, and it is bounded by the
\emph{Gribov horizon}, where $M$ develops a zero eigenvalue. In
principle, the functional integral in minimal Coulomb gauge should
be restricted to a subspace of the Gribov region in which the
gauge fields are global maxima of $R$; this subspace is known as
the ``Fundamental Modular Region."  Part of the boundary of the
fundamental modular region lies on the Gribov horizon.

   The dimension of lattice configuration space is of course very
large, on the order of the number of lattice sites, and it has been
proven that, in contrast to an abelian theory, the Gribov region is
bounded and convex (cf.\ \cite{semenov} and section \ref{convex}
below).  A suggestive analogy is that of a sphere in $N$-dimensions,
which has a volume element proportional to $r^{N-1} dr$, so that most
of the volume is concentrated close to the surface of the sphere. By
this analogy, it is reasonable to suppose that the volume of the
Gribov region is concentrated near the Gribov horizon.  If that is so,
then typical configurations generated by lattice Monte Carlo, fixed to
Coulomb gauge by standard over-relaxation techniques, will also lie
very close to the Gribov horizon, and the F-P operator for such
configurations will acquire near-zero eigenvalues.
This is not enough by itself to ensure
confinement; even the laplacian operator will have a spectrum of
near-zero modes at large lattice volumes.  The conjecture is that
typical configurations near the Gribov horizon may also have enhanced
values for $\r(\l)$ and $F(\l)$ (compared to the perturbative
expressions) at $\l\ra 0$, such that the confinement condition
\rf{condition} is satisfied.  Our task is to check, numerically,
whether this enhancement exists or not. \\

\subsection{Observables}\label{observables}

     We apply the ARPACK routine \cite{ARPACK}, which employs a
variant of the Arnoldi procedure for sparse matrices, to evaluate
the lowest 200 eigenvectors and corresponding eigenvalues of the
F-P matrix $M^{ab}_{xy}$ in eq.\ \rf{M}, for configurations generated
by lattice Monte Carlo.  The first three
eigenvalues are zero for SU(2) gauge theory, regardless of the
lattice configuration, due to the fact that the eigenvector
equation \rf{eigen} is trivially satisfied by three linearly
independent, spatially constant eigenvectors
\begin{equation}
        \p^{a(n)}_x = {1\over \sqrt{V_3}} \d_{an} ~,~~ n=1,2,3
\label{trivial}
\end{equation}
with zero eigenvalue.

    The existence of these trivial zero modes is related to the fact
that physical states with non-zero total color charge cannot exist in
a finite volume with periodic boundary conditions. This is true even
for an abelian theory, and the reason is simple: the Gauss law cannot
be satisfied for total non-zero charge in a finite periodic lattice.
In such cases, the electric flux lines diverging from point sources
have nowhere to end.  This means that the F-P operator (or the
laplacian, in an abelian theory) is non-invertible on a periodic
lattice.  It is precisely the existence of the trivial zero modes
which makes the F-P operator non-invertible; there is no such
difficulty in an infinite volume.  In order to extrapolate our results
on finite volumes to infinite volume, which allows non-zero total
charge, there are two possibilities.  First, we could get rid of zero
modes by imposing non-periodic (e.g.\ Dirichlet) boundary conditions
on the finite lattice.  Second, we could perform our Monte Carlo
simulations on finite periodic lattices as usual, but drop the trivial
zero modes before extrapolating our results to infinite volume.  In
this article we choose the second approach, and exclude the trivial
zero modes from all sums over eigenstates.

    The average eigenvalue density $\rho(\l)$ is
obtained from the remaining 197 eigenvalues in each thermalized
configuration (there are $L$ such configurations in a given $L^4$
lattice, one at each time-slice).  The range of eigenvalues is
divided into a large number of subintervals, and eigenvalues are
binned in each subinterval to determine the expected number
$N(\l,\l+\D \l)$ of eigenvalues per configuration
falling into each bin.  Substituting this value into the
definition \rf{density} of the normalized density of states, we obtain an
approximation to $\rho(\l)$ for $\l$ values in the
middle of each subinterval.  We also compute the expectation value
of the $n$-th eigenvalue and corresponding quantity $F(\l_n)$
\begin{equation}
        \langle \l_n \rangle ~~\mbox{and}~~ \langle F_n \rangle \equiv \langle
        \vec{\p}^{(n)*} \cdot (-\nabla^2)
         \vec{\p}^{(n)} \rangle
\end{equation}
for the $n=4-200$ non-trivial eigenvectors.  Our plots of $F(\l)$ vs.\
$\l$, shown below, are obtained by plotting $\langle F(\l_n) \rangle$
vs.\ $\langle \l_n \rangle$.   Finally, we
calculate the average contribution of low-lying eigenstates with
$\l_n < 0.15$ to the energy $\E$ of unscreened color charge:
\begin{equation}
       \e = {1\over 3V_3}  \sum_{n>3}^{\l_n<0.15}
               \left\langle {F_n \over \l_n^2} \right\rangle .
\label{e}
\end{equation}
For our purposes, the precise value of the upper limit in the sum is
not too important.  We have chosen the upper limit $\l=0.15$ in order
that the 200 lowest eigenvalues, on each lattice volume we have
studied, include the range $0 < \l \le 0.15$.

\subsection{Center projection and vortex removal}

    Not every configuration on the Gribov horizon satisfies the
confinement condition \rf{condition}. For example, \emph{any} purely
abelian configuration in a non-abelian theory lies on the Gribov
horizon (after gauge transformation to Coulomb gauge~\cite{Us}) and
therefore has a (non-trivial) zero eigenvalue, but not all such
configurations will disorder Wilson loops, or lead to an F-P
eigenvalue spectrum satisfying eq.\ \rf{condition}.  The center vortex
theory of confinement holds that a particular class of field
configurations dominates the vacuum state at large scales, and is
responsible for the linear rise of the static quark potential.  If so,
then these same configurations should be responsible for the pileup of
F-P eigenvalues near $\l=0$, resulting in the infinite energy of an
isolated color charge. This is the connection which we think must
exist between the center vortex and Gribov horizon confinement
scenarios.

     To investigate this connection, we apply standard methods to
factor a lattice configuration into its vortex and non-vortex
content.  This is done by first fixing to direct maximal center
gauge, which is the Landau gauge condition in the adjoint
representation
\begin{equation}
          R = \sum_{x} \sum_\m \mbox{Tr}[U_\m(x)]^2  ~~ \mbox{is maximum}
\end{equation}
using an over-relaxation technique. The lattice
configuration is factored into
\begin{equation}
          U_\m(x) = Z_\m(x) \tU_\m(x)
\end{equation}
where
\begin{equation}
         Z_\m(x) = \mbox{sign}\{\mbox{Tr}[U_\m(x)]\}
\end{equation}
is the center-projected configuration, and $\tU_\m$ is the
\emph{``vortex-removed''} configuration. The center-projected (thin vortex)
configuration $Z_\m(x)$ carries the fluctuations which give rise to an area
law for Wilson loops.  The asymptotic string tension of the
vortex-removed configuration $\tU_\m(x)$ vanishes, as does its chiral
condensate and topological charge.  The numerical evidence
supporting these statements is reviewed in ref.\ \cite{review}.

     In our procedure, each thermalized lattice configuration is
transformed to direct maximal center gauge and factored as above into
a center-projected configuration $Z_\m(x)$, and a vortex-removed
configuration $\tU_\m(x)$.  These are then transformed separately into
Coulomb gauge. Of course, any center-projected configuration $Z_\m(x)$, with
links $=\pm I$, trivially fulfills the Coulomb gauge condition
\begin{equation}
       \sum_k \mbox{Tr}\Bigl[\s^a(U_k(x) + U^\dg_k(x-\widehat{k}))\Bigr] = 0,
\end{equation}
but in general such configurations are far from the minimal Coulomb gauge, and
are normally not even in the Gribov region.  So in practice we perform
a random gauge transformation on $Z_\m(x)$, and then fix to a gauge copy
in the Gribov region by the usual over-relaxation method.  We will refer
to such copies as \emph{``vortex-only''} configurations.
Applying the Arnoldi algorithm to calculate the F-P eigenvectors and
eigenvalues, we compute observables $\e,~\r(\l),~\langle
F_n \rangle,~\langle \l_n \rangle$ for both the vortex-only and
vortex-removed configurations.

     Any purely center configuration lies on the Gribov horizon.  By
``center configuration" we mean a lattice all of whose link variables
can be transformed, by some gauge transformation, to center elements
of the gauge group, and the ``vortex-only'' configurations are center
configurations in this sense.  It was shown in ref.\ \cite{Us} that
for the SU(2) gauge group, such configurations have in general three
non-trivial F-P zero modes, in addition to the three trivial,
spatially constant zero modes \rf{trivial}.\footnote{In some special
cases of high symmetry there may be fewer non-trivial zero modes.}  In
computing $\e$ for the vortex-only configurations, we therefore
exclude the first six eigenvalues.

\subsection{Results}

    Most of our simulations have been carried out at $\b=2.1$.  This is not
such a weak coupling, but it allows us to use modestly sized
lattices whose volumes are nonetheless quite large compared to the
confinement scale, and to study the volume dependence.  The F-P
observables are calculated in the full configurations, the
thin-vortex configurations, and the vortex-removed configurations,
each of which has been transformed to Coulomb gauge.

     Figures \ref{r} and \ref{f} show our results for $\r(\l)$
and $F(\l)$
for the full configurations, on a variety of lattice volumes
ranging from $8^4$ to $20^4$ (to reduce symbol overlap near $\l=0$,
we do not display the entire set of data points in $F(\l)$).
The apparent sharp ``bend" in
$\r(\l)$ near $\l=0$ becomes increasingly sharp, and happens ever
nearer $\l=0$, as the lattice volume increase. The impression
these graphs convey is that in the limit of infinite volume,
both $\rho(\l)$ and $F(\l)$ go to positive constants as $\l \ra 0$.
However, for both $\rho(\l)$ and $F(\l)$ we cannot
exclude the possibility that the curves behave like $\l^p,~\l^q$ near
$\l=0$, with $p,q$ small powers.
If we assume that the low-lying eigenvalue distribution scales with
the total number of eigenvalues ($3L^3$)
in the manner suggested by random matrix theory, then it
is possible to deduce, from the probability distribution of the
lowest non-zero eigenvalue, the power dependence of $\rho(\l)$ near $\l=0$.
This analysis is carried out in Appendix \ref{A}, and gives
us the estimates
\begin{equation}
          \rho(\l) \sim \l^{0.25} ~,~
            F(\l)  \sim \l^{0.38}  ~~~\mbox{full configurations}
\end{equation}
at small $\l$ and large volume, with perhaps a 20\% error in the exponents.
With this behavior the
Coulomb confinement condition is satisfied, and the Coulomb self-energy
is infrared divergent.

\FIGURE[b]{
\centerline{\includegraphics[width=8truecm]{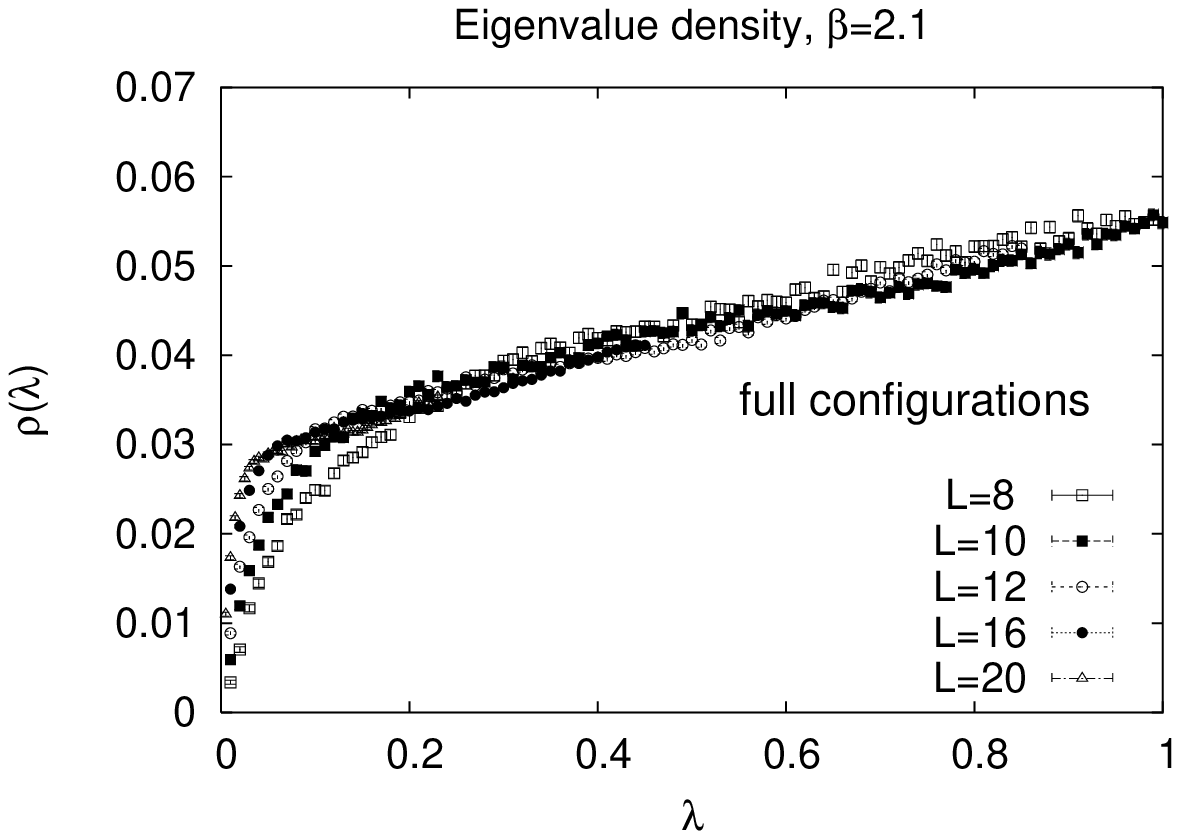}}
\caption{The F-P eigenvalue density at $\b=2.1$, on $8^4-20^4$
lattice volumes.} \label{r}}

\FIGURE[t]{
\centerline{\includegraphics[width=8truecm]{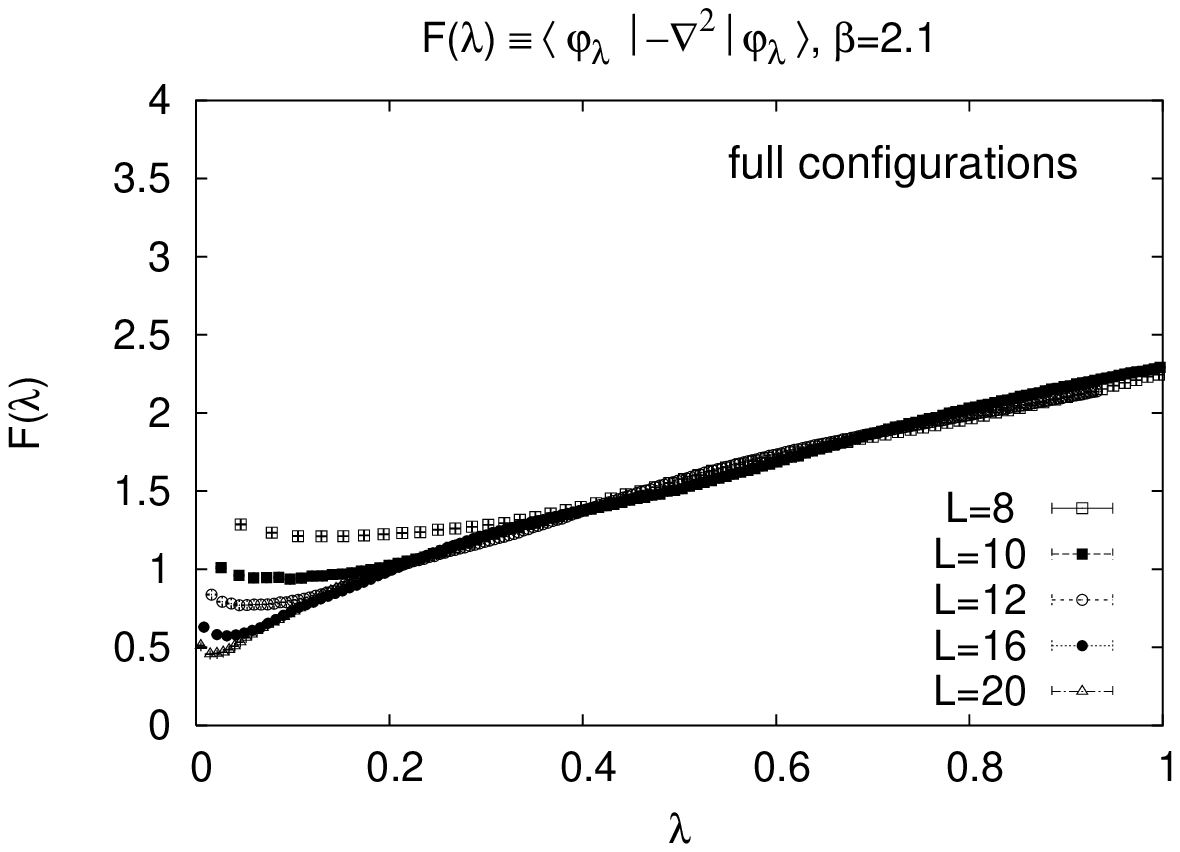}}
\caption{$F(\l)$, the diagonal matrix element of $(-\nabla^2)$ in
F-P eigenstates, plotted vs.\ F-P eigenvalue.} \label{f}}

    In Fig.\ \ref{s1full} we plot $\e$ vs.\ lattice extension $L$,
together with a best straight-line fit through the points at
$L=10-20$.  The cut-off energy $\e$ rises with $L$, and this rise
is consistent with linear, although the data is not really good enough to
determine the precise $L$-dependence.

\FIGURE[b]{
\centerline{\includegraphics[width=8truecm]{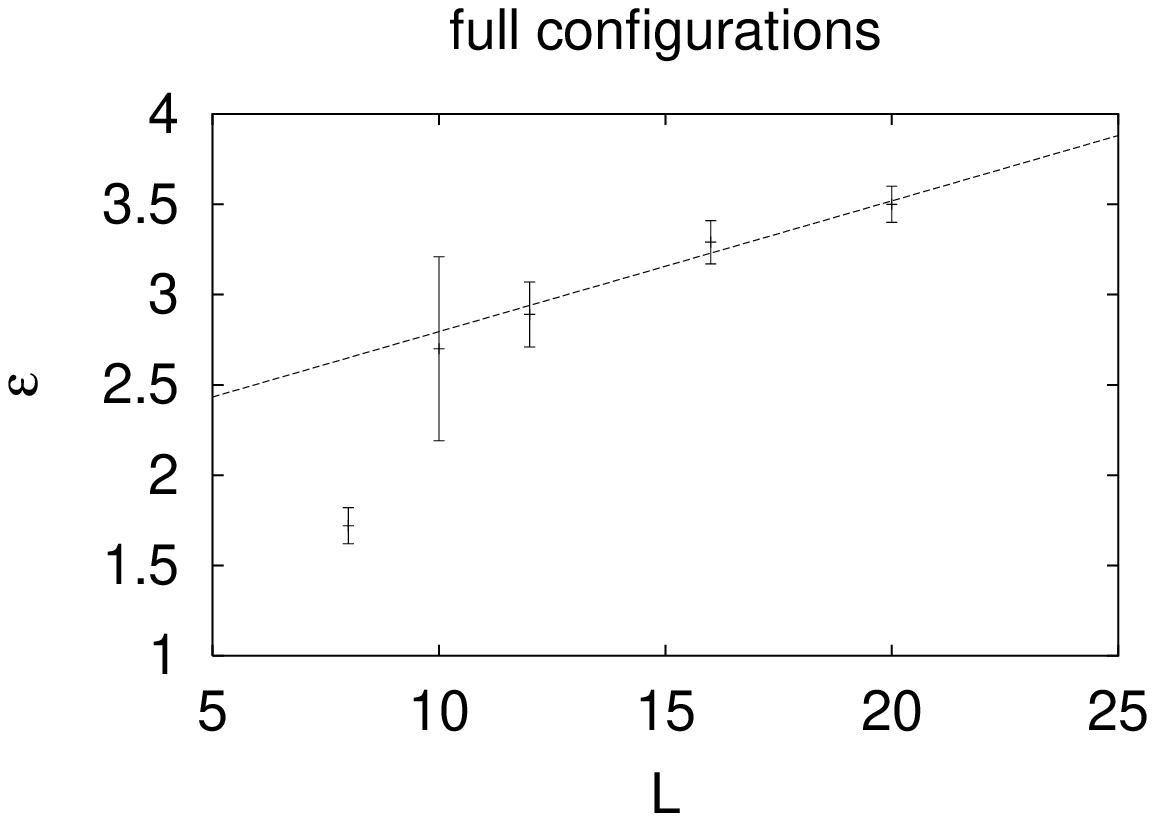}}
\caption{$\e$ vs.\ lattice extension $L$, for full, unprojected
lattice configurations.  The straight line is a linear fit through
points at $L=10-20$.} \label{s1full}}

    One might wonder how it is possible that
a pair of quark-antiquark charges, in a global color singlet
combination, can have a finite total Coulomb energy in an infinite
volume, given that each charge has a divergent self-energy. This
question was addressed in ref.\ \cite{Us}, which computed
Coulomb energies from timelike link correlators.  The answer is
that both the quark self-energies, and the energy due to
instantaneous one-gluon exchange between separated sources,
contain an infrared divergent constant.  It can be shown that for
global color singlets these constants precisely cancel, while in
non-singlets the self-energy is not entirely cancelled, and the
total energy is infrared divergent.

    Next we consider the F-P observables for the ``vortex-only" configurations,
consisting of thin vortex configurations (in Coulomb gauge) which
were extracted from thermalized lattices as described above.  Our
data for $\r(\l)$ and $F(\l)$ at the
same range ($8^4-20^4$) of lattice volumes is displayed in Figs.\
\ref{rcp} and \ref{fcp}.
The same qualitative features seen for
the full configurations, e.g.\ the sharp bend in the eigenvalue
density near $\l=0$, becoming sharper with increasing volume, are
present in the vortex-only data as well, and if anything are more
pronounced. From the graphs, it would appear that
\begin{equation}
     \rho(0) \approx 0.06  ~~,~~ F(0) \approx 1.0 ~,~~~~ \mbox{vortex-only.}
\label{vo}
\end{equation}
This non-zero limit for $\rho(\l),~F(\l)$ at $\l\ra 0$
is supported by an analysis of the low-eigenvalue universal scaling behavior
as a function of $L$, which is reported in Appendix \ref{A}.
Once again, the confinement criterion \rf{condition} is obviously satisfied.

\FIGURE[t]{
\centerline{\includegraphics[width=8truecm]{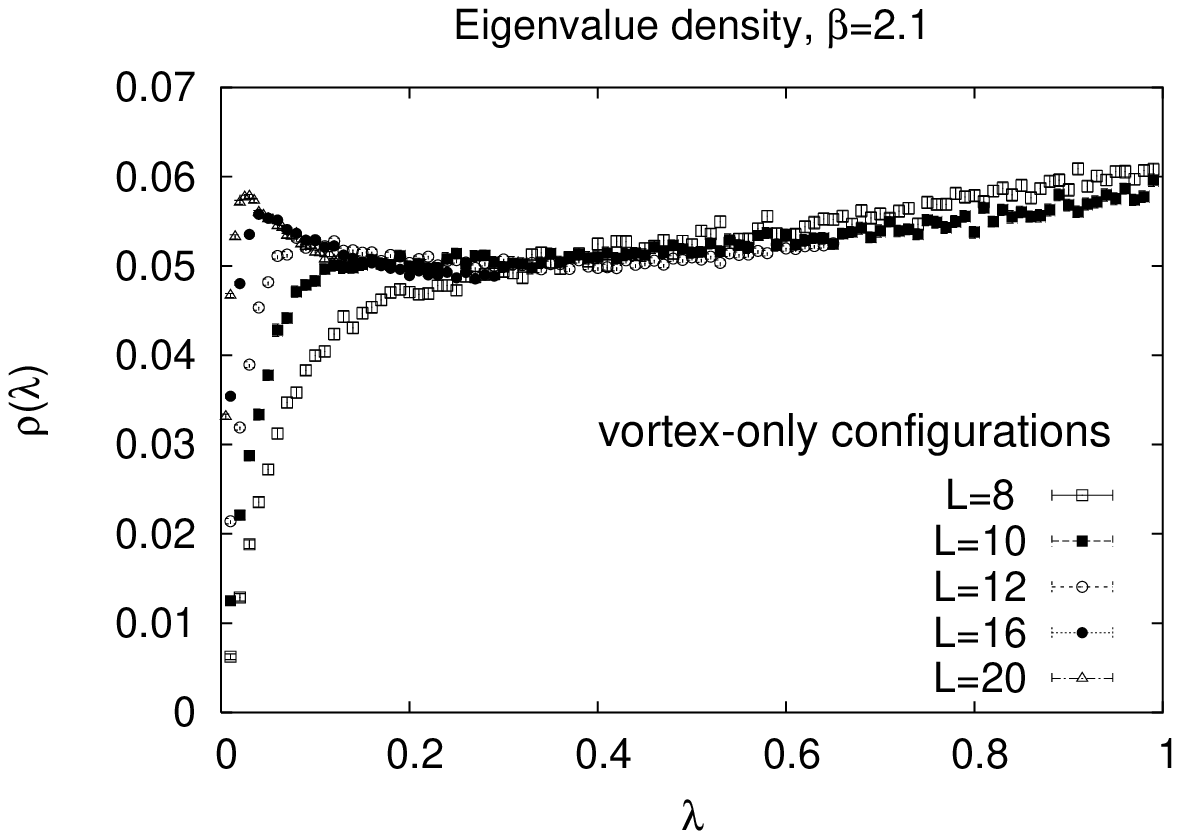}}
\caption{F-P eigenvalue density in vortex-only configurations.}
\label{rcp}}

\FIGURE[b]{
\centerline{\includegraphics[width=8truecm]{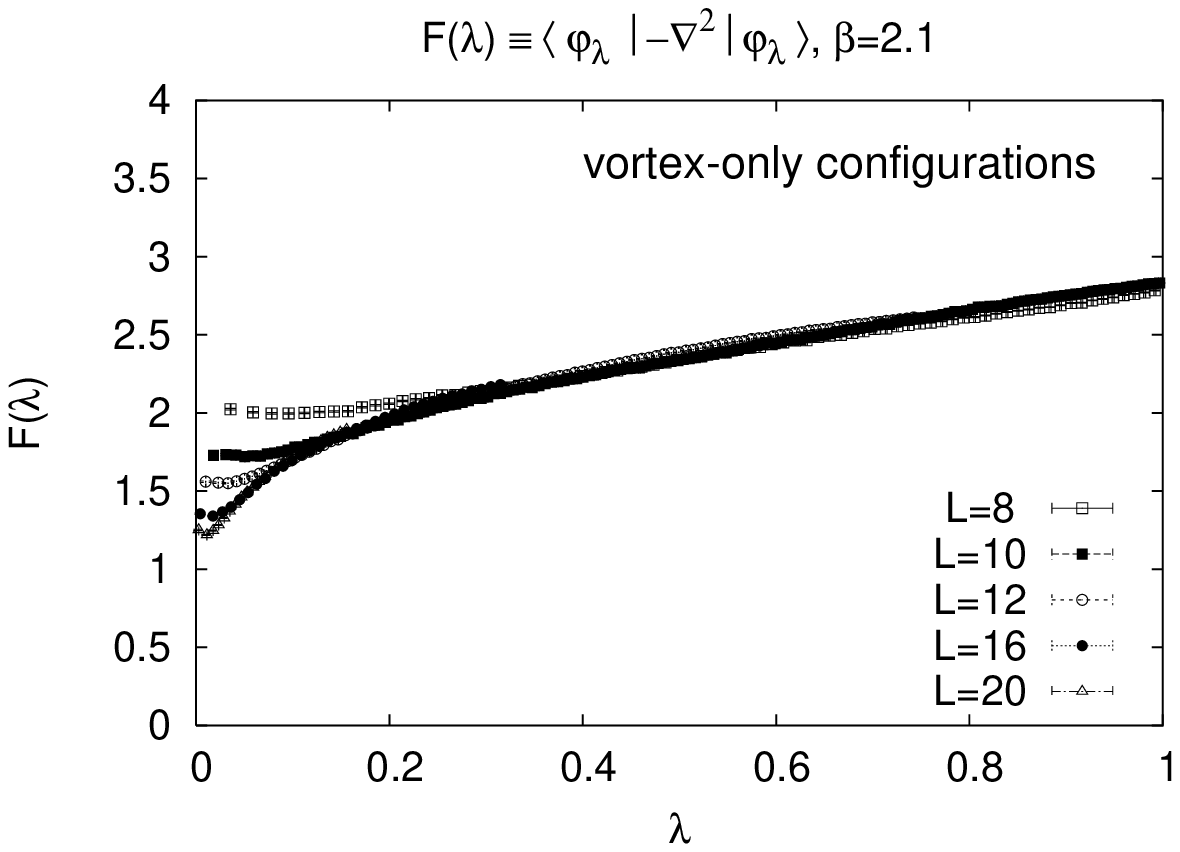}}
\caption{$F(\l)$, the diagonal matrix element of $(-\nabla^2)$ in
F-P eigenstates, for vortex-only configurations.} \label{fcp}}

    Figure \ref{s1cp} shows our data for $\e(L)$, again with a linear
fit through the data points at $L=10-20$, although the linear dependence
is not really established.

\FIGURE[t]{
\centerline{\includegraphics[width=8truecm]{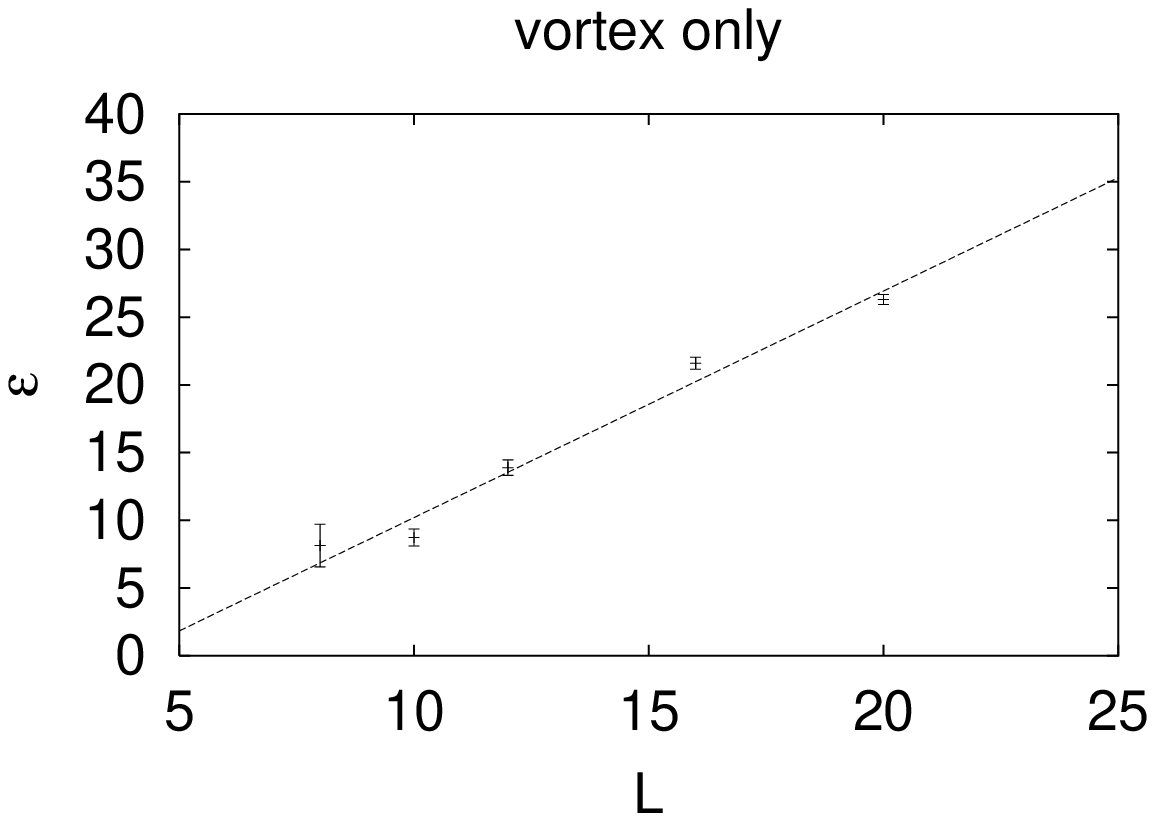}}
\caption{$\e$ vs.\ lattice extension $L$, for the vortex-only
lattice configurations.  The straight line is a linear fit through
points at $L=10-20$.} \label{s1cp}}

    Finally, we consider the F-P observables of the vortex-removed
configurations $\tU$ transformed to Coulomb gauge.  Our results are
shown in Fig.\ \ref{rnv} for $\r(\l)$, and Fig.\ \ref{fnv} for
$F(\l)$. The behavior of these observables is strikingly different,
in the vortex-removed configurations, from what is seen in the full
and vortex-only configurations.  A graph of the eigenvalue density,
at each lattice volume, shows a set of distinct peaks, while the
data for $F(\l)$ is organized into bands, with a slight gap between
each band.  Closer inspection shows that eigenvalue interval
associated with each band in $F(\l)$ precisely matches the
eigenvalue interval of one of the peaks in $\r(\l)$.

\TABLE[r]{ \centering
\begin{tabular}{cccc}
   \hline
   $k$ & $N_k$ & $n_i$ & $\l_k$ \\
   \hline
    1 &  3  &   $(0,0,0)$ &  0 \\
    2 & 18  &   $(1,0,0)$ & $4\sin^2(\pi/L)$ \\
    3 & 36  &   $(1,1,0)$ & $8\sin^2(\pi/L)$ \\
    4 & 24  &   $(1,1,1)$ & $12\sin^2(\pi/L)$ \\
    5 & 18  &   $(2,0,0)$ & $4\sin^2(2\pi/L)$ \\
    \hline
\end{tabular}
\caption{Eigenvalues $\l_k$ of the zero-field lattice F-P operator
$-\d^{ab}\nabla^2$, and their degeneracies $N_k$.}\label{freeth}}

    In order to understand these features, we consider the eigenvalue
density of the F-P operator $M^{ab}_{xy}=\d^{ab}(-\nabla^2)_{xy}$
appropriate to an abelian theory (or a non-abelian theory at zero-th
order in the coupling).  Although we can readily derive the result
$\rho(\l) \sim \sqrt{\l}$ \rf{zeroth} at infinite volume, this
result is slightly misleading at finite volume, where the eigenvalue
density is actually a sum of delta-functions \begin{equation}
       \rho(\l) = {1\over 3V_3} \sum_k N_k \d(\l-\l_k).
\label{peaks} \end{equation} In the sum, the index $k$ labels the
\emph{distinct} eigenvalues of the lattice laplacian $-\nabla^2$,
and $N_k$ is the degeneracy of $\l_k$.  Explicitly, to each distinct
eigenvalue $\l_k$ on an $L^4$ lattice, there is a set of integers
$n_{1-3}$ such that
\begin{equation}
        \l_k = 4\sum_{i=1}^3 \sin^2\left[{\pi n_i\over L}\right].
\end{equation}
The first few values of $\l_k$, and their degeneracies, are listed
in Table \ref{freeth}.

\FIGURE[t]{
\centerline{\includegraphics[width=8truecm]{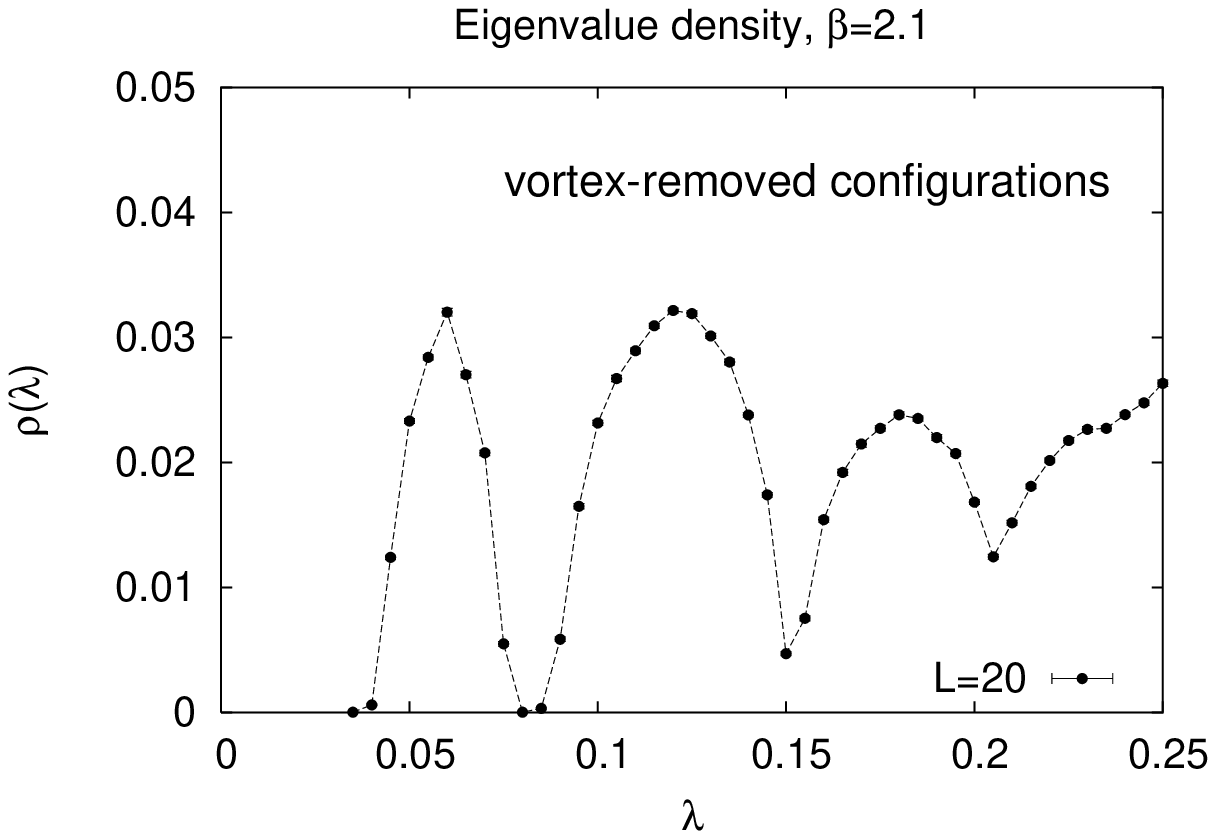}}
\caption{F-P eigenvalue densities for vortex-removed configurations,
on a $20^4$ lattice volume.} \label{rnv}}

\FIGURE[b]{
\centerline{\includegraphics[width=15truecm]{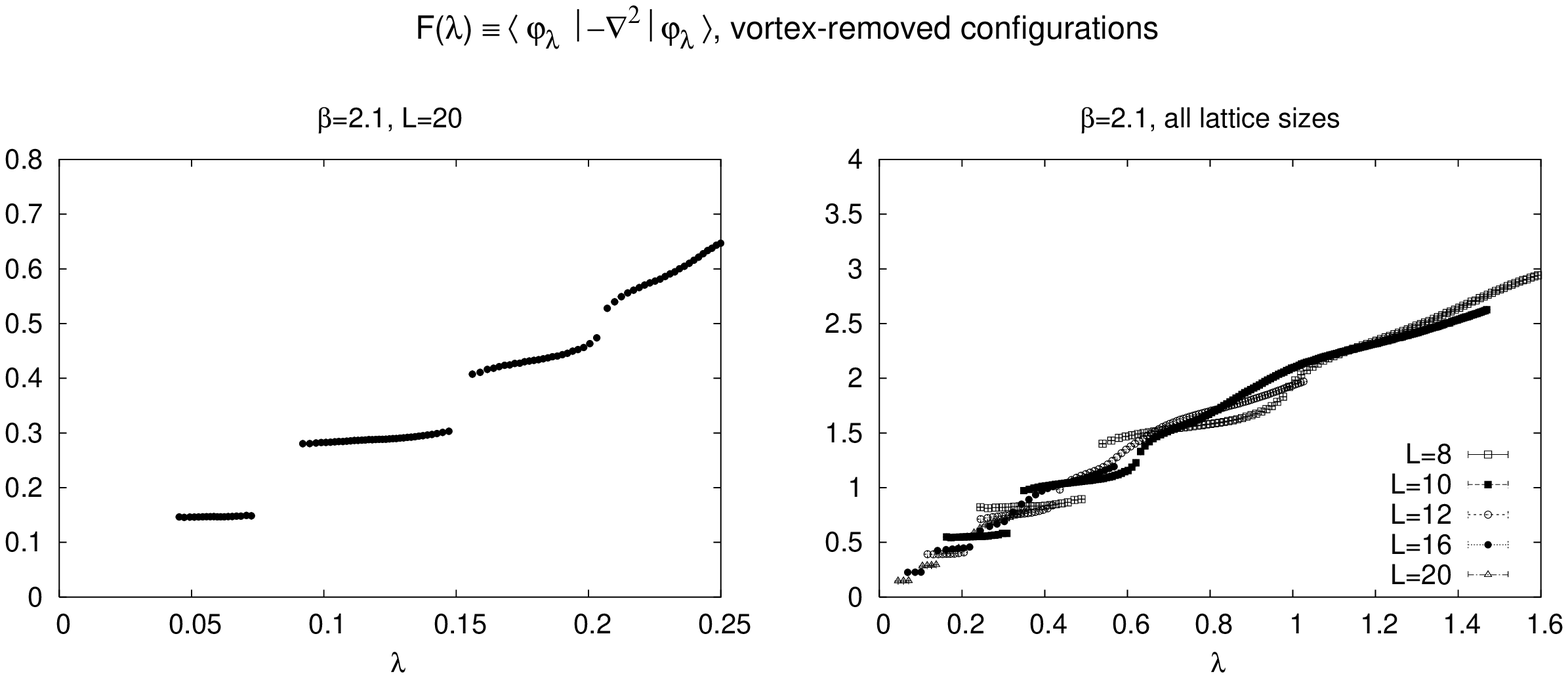}}
\caption{$F(\l)$ vs.\ $\l$ in the vortex-removed configurations.  We
display results for the $20^4$ volume alone (left), and a variety of
lattice volumes (right).} \label{fnv}}

     We have compared these degeneracies $N_k$, of the zeroth-order
F-P eigenvalues, with the number of eigenvalues per lattice
configuration found inside the $k$-th ``peak" of
$\r(\l)$, and $k$-th ``band" of $F(\l)$. We find
there is a precise match.  This leads to a simple interpretation:
the vortex-removed configuration $\tU_\m$ can be treated as a
small perturbation of the zero-field limit $U_\m=I$.  This
perturbation lifts the degeneracy of the $\l_k$, spreading the
degenerate eigenvalues into the bands of finite width in $\l$.  At
least for small $k$, these bands do not overlap. Likewise, the
perturbation broadens the infinitely narrow $\d$-function peaks in
the density of eigenstates, eq.\ \rf{peaks}, into the peaks of
finite width seen in Fig.\ \ref{rnv}.

    Because both the density of eigenvalues and the data for
$F(\l)$ seem to be only a perturbation of the
corresponding zero-field results, it appears to be most unlikely
that the no-vortex configurations lead to a divergent Coulomb
self-energy. In fact, we find that the low-lying eigenvalue spectrum
scales with lattice extension $L$ as
\begin{equation}
        \langle \l_n \rangle \sim {1\over L^2} ,
\end{equation}
just as in the zero-field (or zero-th order) case.
%Finite-size
%scaling arguments (Appendix A) then suggest that $\r(\l) \sim \sqrt{\l}$.
%Likewise, from Fig.\ \ref{fnv} it would appear that $F(\l) \sim \l$.
%These are also the zero-th order behaviors, and it is clear that
%the product $\r(\l) F(\l)$ does not satisfy the confinement condition
%\rf{condition}.
We have not plotted $\e(L)$ for the vortex removed
case, because for the smallest lattice volume no eigenvalues actually
lie in the given range $0<\l<\l_{min}$, and even at $L=10,12$ only a
few of the lowest eigenvalues are in this range.  We do note, however,
that the values of $\e$ on the large lattices are roughly two orders
of magnitude smaller than the corresponding values for the full
configurations.

   Vortex removal is in some sense a minimal disturbance of the
original lattice configuration.  The procedure only changes field
strengths at the location of P-vortex plaquettes, and the fraction
of P-plaquettes out of the total number of plaquettes on the
lattice dives exponentially to zero with increasing $\b$.
Nevertheless, this modest change of the lattice configuration is
known to set the string tension and topological charge to zero,
and to remove chiral symmetry breaking.  At $\b=2.1$, we have
found that vortex removal also drastically affects the density of
low-lying eigenvalues, and one sees, in the multi-peak structure
of $\r(\l)$, the remnants of the delta-function peaks of the free
theory.  We have used this comparatively low value of $\b$ so that
we may probe large physical distances, as compared to the distance
scale set by the string tension, on rather modestly sized (up to
$20^4$) lattices.  This allows us carry out the finite volume
scaling analysis reported in Appendix A.

    However, using such a small $\b$ has a price, in terms of
our confidence in the effects of vortex removal.  At $\b=2.1$
roughly 17\% of all plaquettes are P-vortex plaquettes (cf.\ Fig.\
20 in ref.\ \cite{review}), and one may object that in this case
vortex removal is not such a small disturbance of the lattice
configuration. Perhaps the drastic effect of vortex removal on the
eigenvalue density is simply an artifact of the substantial number
of plaquettes modified.

\begin{figure}[htp]
 \centering
    \subfigure[$\r(\l),~\b=2.3$]{\label{a1}
         \includegraphics[scale=0.5]{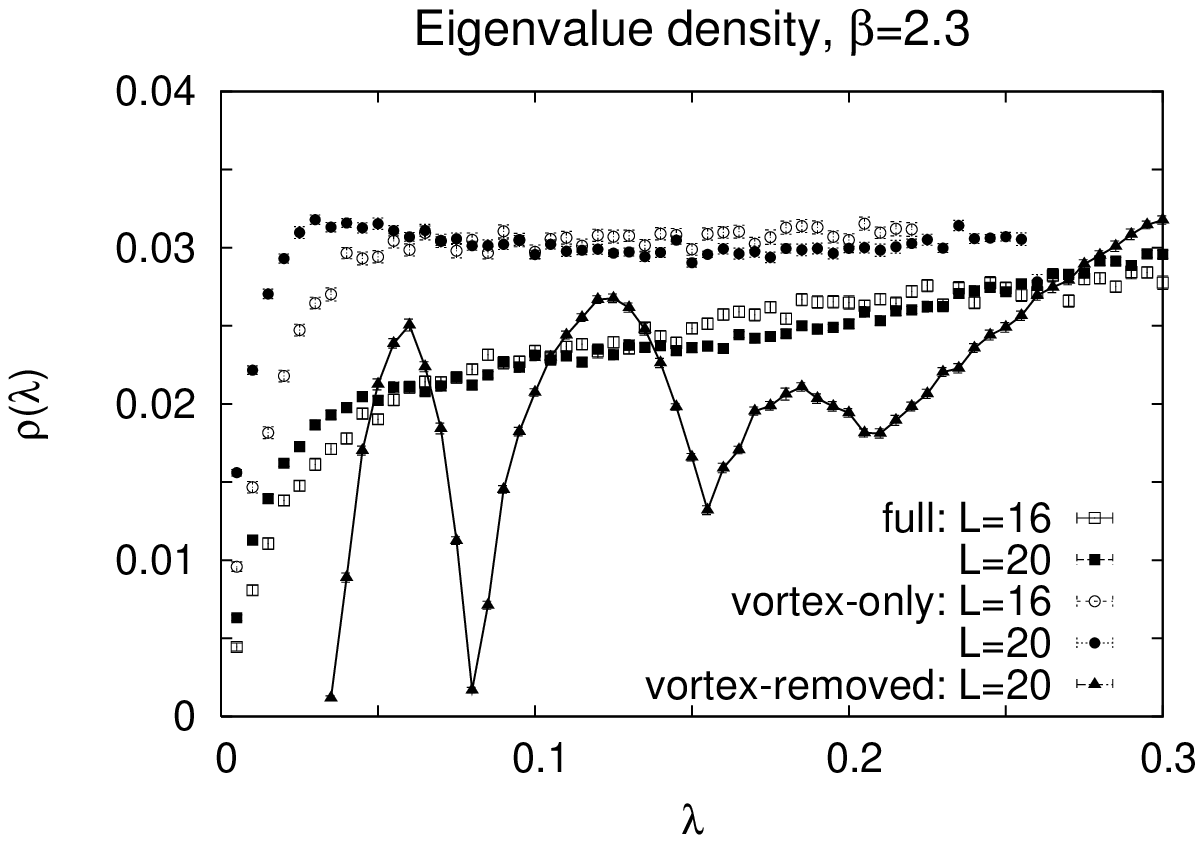}}
          \qquad \qquad
    \subfigure[$\r(\l),~\b=2.4$]{\label{b1}
         \includegraphics[scale=0.5]{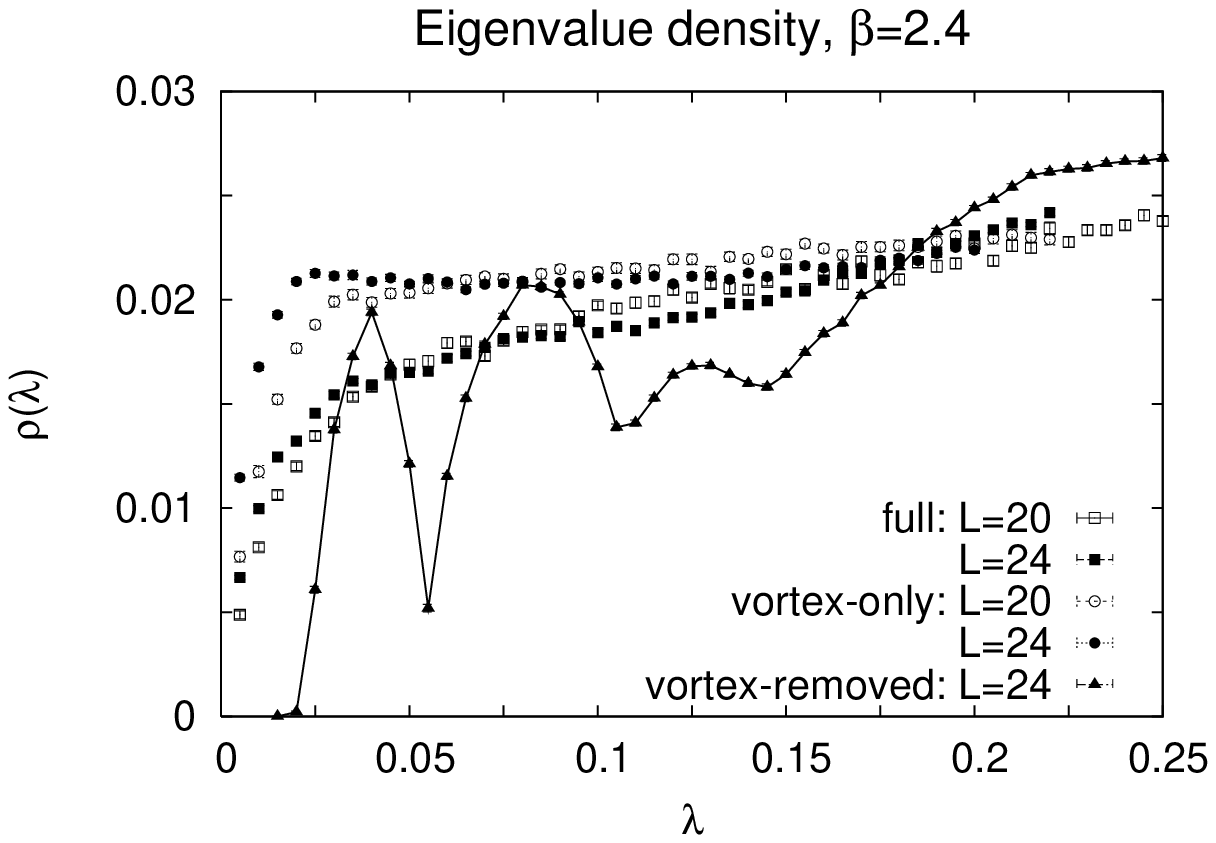}}
%  \caption{Eigenvalue densities at $\b=2.3,~2.4$.}
\centering
    \subfigure[$F(\l),~\b=2.3$]{\label{c1}
         \includegraphics[scale=0.5]{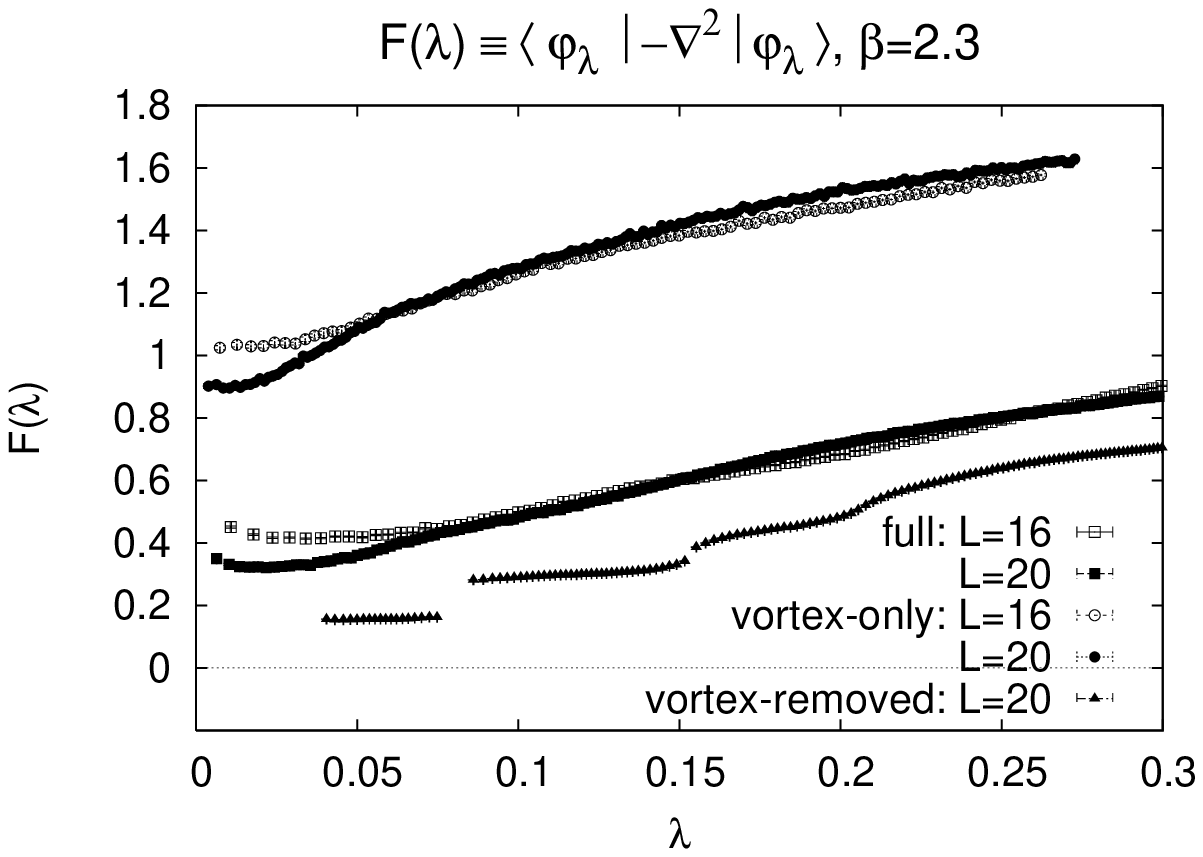}}
          \qquad \qquad
    \subfigure[$F(\l),~\b=2.4$]{\label{d1}
         \includegraphics[scale=0.5]{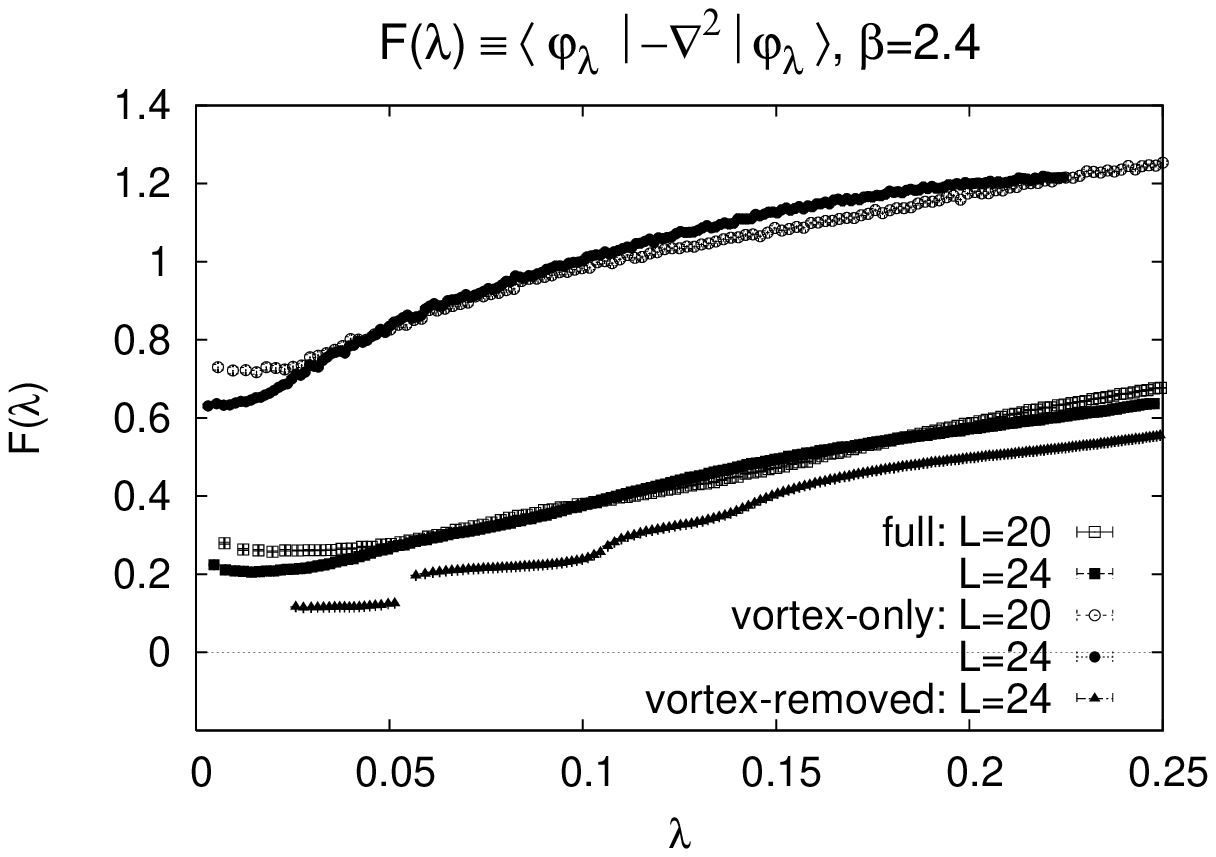}}
  \caption{$\r(\l),~F(\l$) at couplings $\b=2.3,~2.4$ for
  full, vortex-only, and vortex-removed configurations.}
  \label{rfbeta}
\end{figure}

    In order to address this concern, we have computed $\r(\l)$ and $F(\l)$
at $\b=2.3$ and $\b=2.4$, where the P-vortex densities have
dropped to around 9\% and 4\%, respectively, of the total number
of plaquettes.  The results are shown in Fig.\ \ref{rfbeta}, where
we display the data for the unmodified, vortex-only, and
vortex-removed configurations on the same plot.  Only two lattice
volumes are shown at each coupling. The effect of vortex removal
is seen to be much the same at these higher $\b$ values as at
$\b=2.1$. Again we see a multi-peak structure in $\r(\l)$, and a
band structure in $F(\l)$ at the low-lying eigenvalues (although
the gap between bands narrows as $\b$ increases).  In each
configuration, we have checked that the number of eigenvalues in
each peak of $\r(\l)$ matches the number of eigenvalues in each
band of $F(\l)$, and that this number is again equal to the
degeneracy of eigenvalues. Therefore, the interpretation proposed
for the no-vortex data at $\b=2.1$ appears to apply equally well
at the higher $\b$ values: these data are simply perturbations of
the zero-field result, for which the eigenvalue density is a sum
of delta-functions.  The data for the vortex-only and unmodified
configurations are also qualitatively similar to the results we
have obtained at the lower $\b=2.1$ value, although the lattice
volumes, in physical units, are considerably smaller than the
corresponding lattice volumes at $\b=2.1$.

    We conclude that it is the vortex content of the thermalized
configurations which is responsible for the enhancement of both
the eigenvalue density and $F(\l)$ near $\l=0$, leading to an
infrared-divergent Coulomb self-energy.

\section{Gauge-Higgs theory}\label{ghiggs}

    Next we consider a theory with the gauge field coupled to a
scalar field of unit modulus in the fundamental representation of the gauge
group.  For the SU(2) gauge group, the lattice
action can be written in the form \cite{Lang}
\begin{equation}
     S = \b \sum_{plaq} \oh \mbox{Tr}[UUU^\dg U^\dg] \non \\
       + \gamma \sum_{x,\m} \oh \mbox{Tr}[\phi^\dg(x) U_\m(x)
\phi(x+\widehat{\m})]
\end{equation}
with $\phi$ an SU(2) group-valued field.  Strictly speaking, this theory
is non-confining for all values of $\b,\g$.  In particular, there is no
thermodynamic phase transition (non-analyticity in the free energy), from the
Higgs phase to a confinement phase; this is the content of a well-known
theorem by Osterwalder and Seiler \cite{OS}, and Fradkin and
Shenker \cite{FS}.

    However, the Osterwalder--Seiler---Fradkin--Shenker
(OS-FS) theorem is not the last word on
phase structure in the gauge-Higgs system.  In fact, there
\emph{is} a symmetry-breaking transition in this theory.  If one
fixes to Coulomb gauge, there is still a remaining freedom to
perform gauge transformations which depend only on the time
coordinate.  It is therefore possible, in an infinite volume, that
this symmetry is broken on any given time-slice (where the remnant
symmetry is global).  The order parameter for the
symmetry-breaking transition is the modulus of the timelike links,
averaged over any time slice
\begin{eqnarray}
          Q &=& \left\langle \sqrt{\mbox{Tr}[\oh V(t)V^\dg(t)]}\right\rangle,
\non \\
          V &=& {1\over L^3}\sum_x U_0(x,t).
\end{eqnarray}
If $Q\ra 0$ as $L\ra \infty$, then the remnant gauge symmetry
is unbroken, the Coulomb potential (as opposed to the static
potential) between quark-antiquark sources rises linearly, and the
energy of an isolated color-charge state of form \rf{charge1} is
infrared divergent. Conversely, if $Q>0$ at infinite volume, then
the remnant symmetry is broken, the Coulomb potential is
asymptotically flat, and the energy of an isolated color-charge state
is finite.

     These matters are explained in some detail in ref.\ \cite{Us},
where we report on a sharp transition between a symmetric
(``confinement-like") phase and a broken (``Higgs") phase of remnant
gauge symmetry. There is no inconsistency with the OS-FS
theorem, which assures analyticity of local, gauge-invariant order
parameters.  The order parameter $Q$, when expressed as a
gauge-invariant observable, is highly non-local.  The transition line
between the symmetric and broken phases is also not the location of a
thermodynamic transition, in the sense of locating non-analyticity in
the free energy.  Rather, it is most likely a Kert\'esz line, of the
sort found in the Ising model at finite external magnetic field, which
identifies a percolation transition \cite{Kertesz}.  In the
gauge-Higgs case, the objects which percolate in the confinement-like
phase, and cease to percolate in the Higgs phase, turn out to be
center vortices \cite{Bertle2} (see also \cite{Kurt1,Kurt2}).

    In the previous section we investigated the effect, on F-P
observables, of removing center vortices from lattice
configurations by hand. The gauge-Higgs system gives us the
opportunity of suppressing the percolation of vortices by simply
adjusting the coupling constants; we can then study the F-P
observables in screened phases with and without percolation. We
note that the ``confinement-like" phase is a screened phase, rather
than a true confinement phase, in that the energy of a static
color charge is not infinite, because it can be screened by the
dynamical Higgs particles. Nevertheless, in this phase the Coulomb
potential is confining, and the Coulomb energy of an isolated
charged state of the form \rf{charge1} (an unscreened charge) is
infrared infinite \cite{Us}.\footnote{The Coulomb potential is
only an upper bound on the static quark potential \cite{Dan}; a
confining Coulomb potential is a necessary but not sufficient
condition for confinement.}

\FIGURE[t]{
\centerline{\includegraphics[width=8truecm]{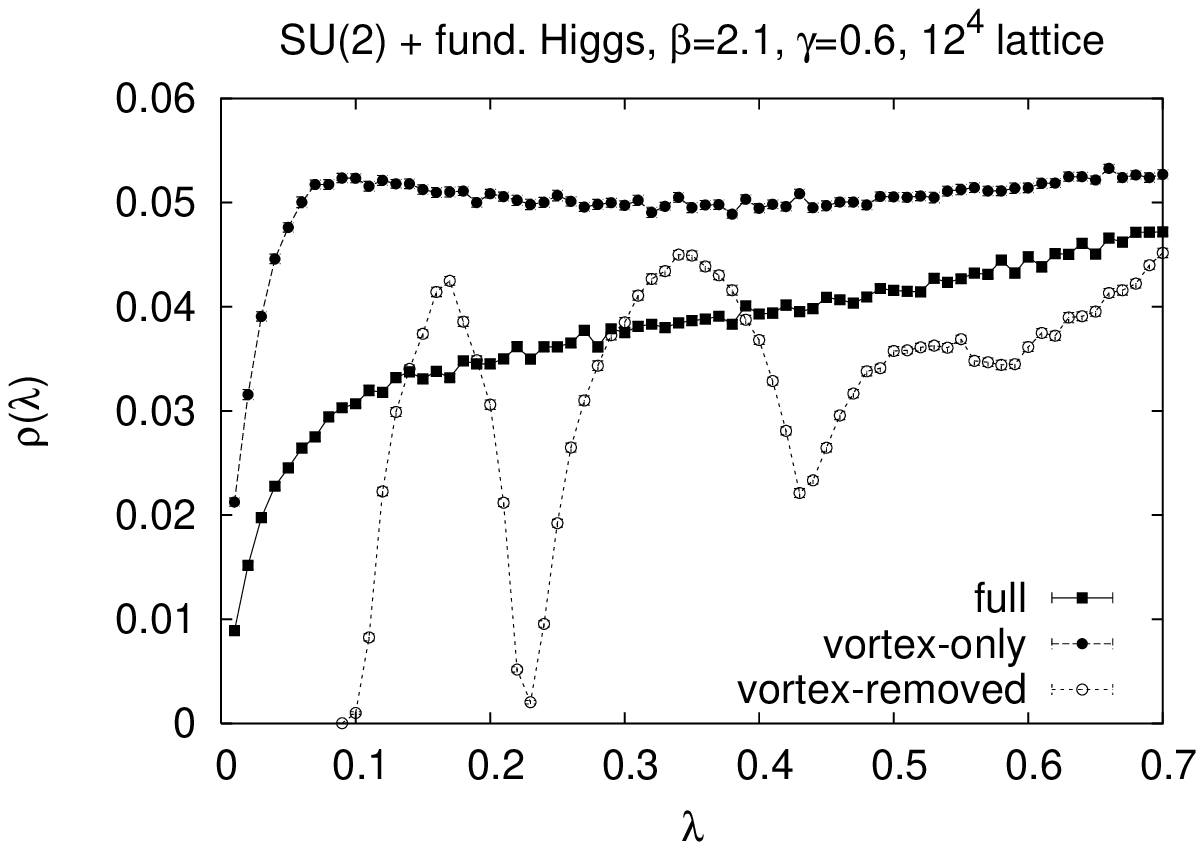}}
\caption{Eigenvalue densities for a gauge-Higgs system in the
``confined", unbroken remnant symmetry phase.  Data for the full,
vortex-only, and vortex-removed configurations are shown, taken on a
$12^4$ lattice.} \label{rgh-con}}

\FIGURE[t]{
\centerline{\includegraphics[width=8truecm]{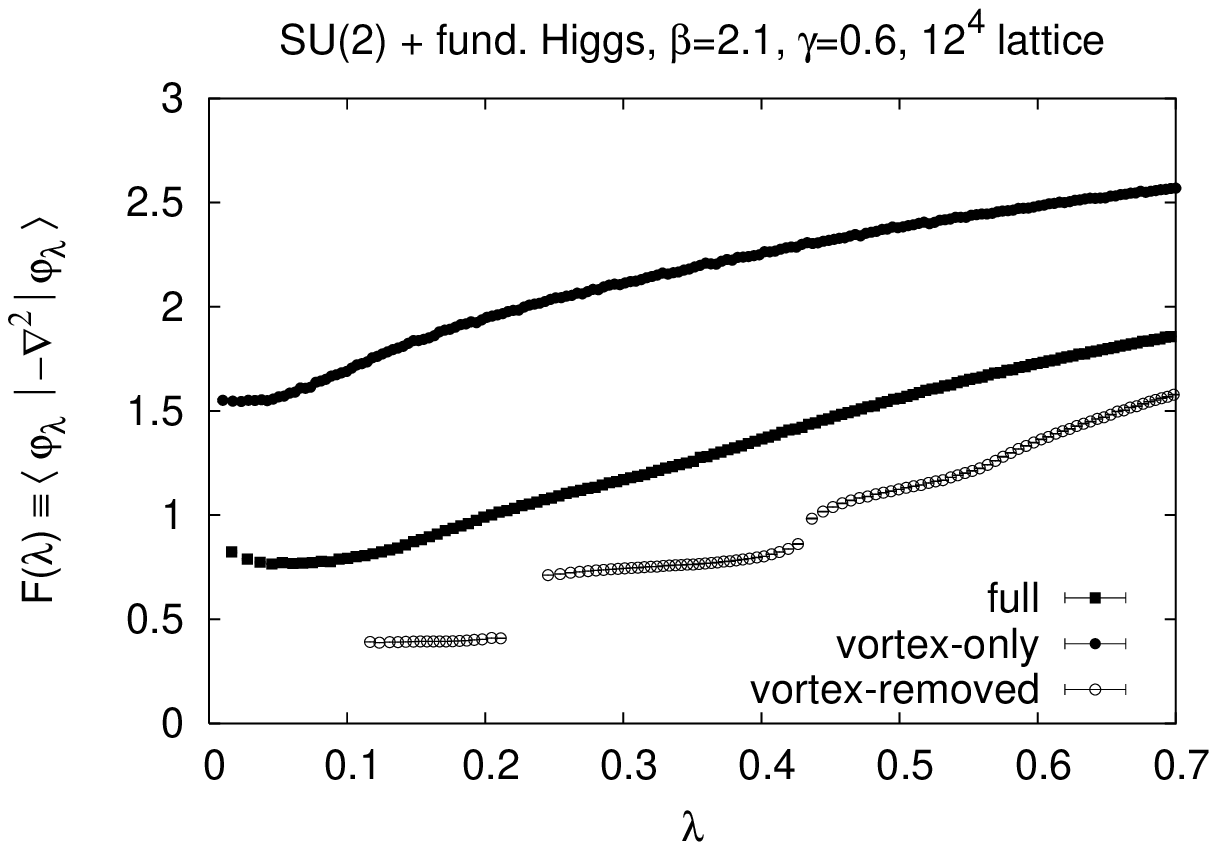}}
\caption{$F(\l)$ for a gauge-Higgs system in the ``confined" phase,
on a $12^4$ lattice.  Data is shown for the full, vortex-only, and
vortex-removed configurations.} \label{fgh-con}}

    In Figs.\ \ref{rgh-con} and \ref{fgh-con} we display the results
for $\r(\l)$ and $F(\l)$ on a
$12^4$ lattice at $\b=2.1$ and $\g=0.6$, which is inside the
symmetric (or ``confinement-like") phase.  Data extracted from the
full lattice, vortex-only, and vortex-removed configurations are
shown on each graph, and these are qualitatively very similar to
what we have already seen in the pure-gauge theory (equivalent to
$\g=0$).  But things change drastically as we move across the
transition into the Higgs phase. The results for $\r(\l)$ and
$F(\l)$ at $\b=2.1$ and $\g=1.2$,
which is inside the Higgs phase, are shown in Figs.\ \ref{rgh-hig}
and \ref{fgh-hig}.  Data, taken on a $12^4$ lattice, is displayed
for the full configurations only. Note that these figures look
almost identical to the corresponding \emph{vortex-removed} data in the
symmetric phase, Figs.\ \ref{rgh-con} and \ref{fgh-con}, which in turn
are very close to the vortex-removed
data obtained in the pure gauge theory at $\b=2.1$,
on a $12^4$ lattice.

\FIGURE[t]{
\centerline{\includegraphics[width=8truecm]{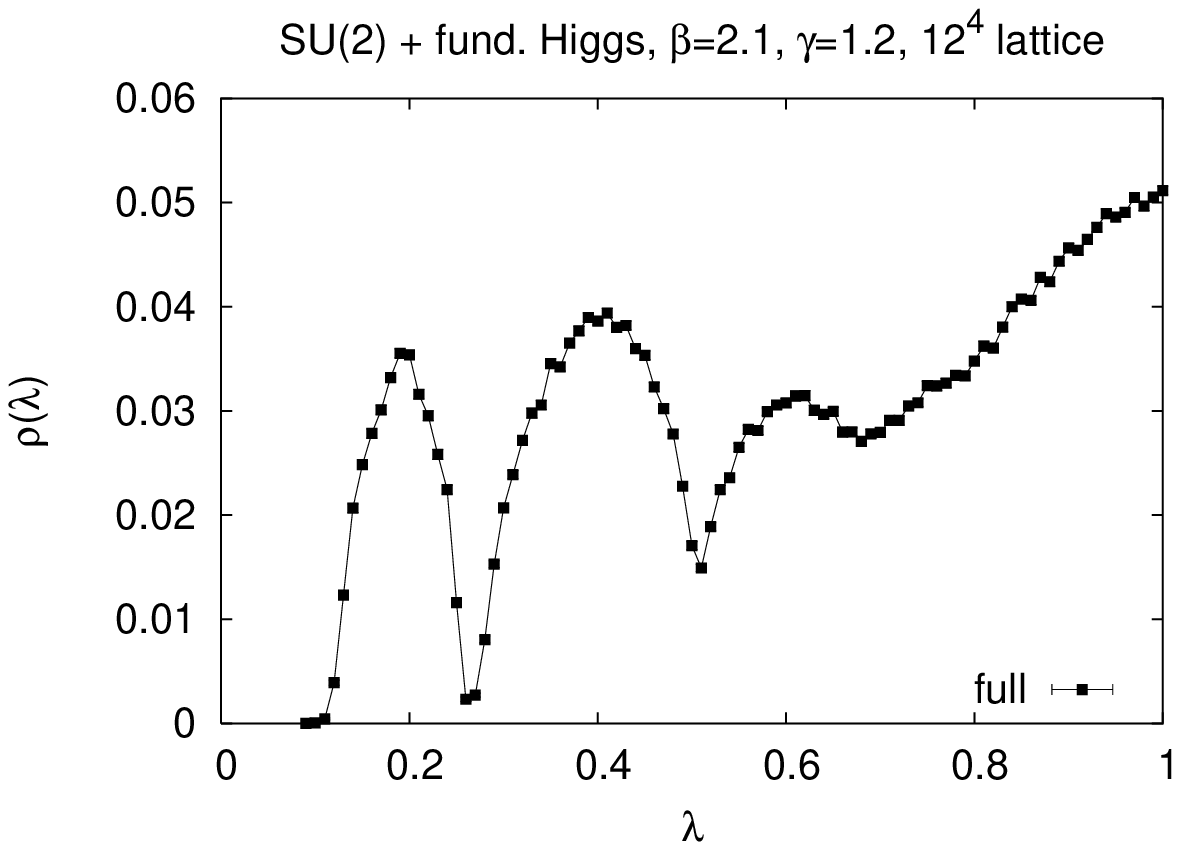}}
\caption{The F-P eigenvalue density in the Higgs (broken remnant
symmetry) phase, for the full configurations.} \label{rgh-hig}}

    This last feature is worth stressing.
From the point of view of the F-P operator, a thermalized
configuration in a pure-gauge theory factors into a piece which
actually does the confining (the vortex-only configuration), and a
piece which closely resembles the lattice of a gauge theory in the
Higgs phase (the vortex-removed configuration).

    Vortex-removal in the Higgs phase does not
affect the data for $\r(\l)$
appreciably.\footnote{The eigenvalue density for vortex-only
configurations shows a set of sharp, narrow peaks, very much like
the sum of delta-functions in the zero-field limit.} It seems safe
to conclude that this behavior of the F-P observables, found in
the Higgs phase, is consistent with an infrared finite Coulomb
self-energy.

\FIGURE[t]{
\centerline{\includegraphics[width=8truecm]{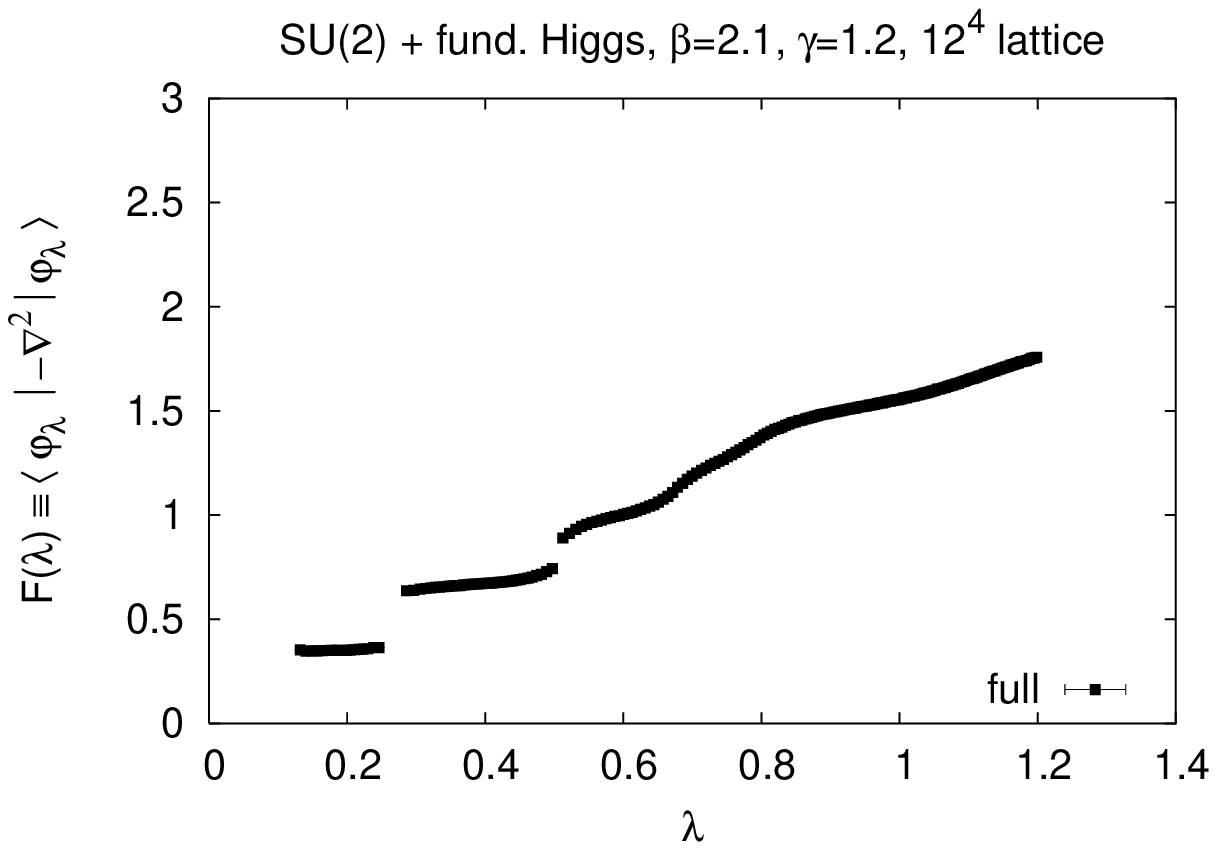}}
\caption{$F(\l)$ in the Higgs phase, for full configurations.}
\label{fgh-hig}}

\section{Eigenvalue density in the deconfined phase}\label{hight}

It was reported in ref.\ \cite{Us} that the instantaneous
color Coulomb potential $V_{coul}(r)$ is linearly rising at large
separation, i.e.\ $V_{coul}(r) \sim \sigma_{coul} r$ ~ where
$\sigma_{coul}$ is a Coulomb string tension, in the
high-temperature deconfined phase of pure gauge theory.
This fact is surprising
at first sight, because the free energy $V(r)$ of a static quark
pair in the deconfined phase is not confining.  But it is not
paradoxical, because $V_{coul}(r)$ is the energy of unscreened
charges, and provides only an upper bound on~$V(r)$ \cite{Dan}.
Confining behavior in $V_{coul}(R)$ is a necessary but
not sufficient condition for confinement, as we have already
noted.

    In fact, the confining behavior of $V_{coul}(R)$ in the deconfinement
phase was to be expected.  The color Coulomb potential is derived
from the non-local term $H_{coul}$ in the Coulomb gauge
Hamiltonian, this term depends on the Faddeev-Popov operator via
the expression $M^{-1}(-\nabla^2)M^{-1}$, and the F-P operator, on
the lattice, depends only on spacelike links at a fixed time. But
we know that even at very high temperatures, spacelike links at
any fixed time form a confining $D=3$ dimensional lattice,
and spacelike Wilson loops have an area-law
falloff just as in the zero-temperature case \cite{spatialWL}.
Since $V_{coul}(R)$ depends only on the three-dimensional lattice,
it is natural that $V_{coul}(R)$ confines at any temperature, and that
remnant gauge symmetry (as opposed to center symmetry) is realized
in the unbroken phase.

   The role of center vortices in producing an area law for spacelike
loops at high temperatures has been discussed in refs.\
\cite{cv-T}. If the confinement property of spacelike links is
eliminated by vortex removal, then by the previous reasoning we
would expect $\s_{coul}$ to vanish as a consequence. This
consequence was also verified in ref.\ \cite{Us}.

Given these results, it may be expected that the color-Coulomb
self-energy $\E_r$ of an isolated static charge is infrared divergent in
the deconfined phase, and that this is associated with an enhancement
of the eigenvalue density $\r(\l)$ of the Faddeev--Popov operator,
and of $F(\l)$, the expectation value of $(- \nabla^2)$ in F-P eigenstates.

To test this we have evaluated $\r(\l)$ and $F(\l)$ at $\b = 2.3$ on
$16^3 \times 2$ and $20^3 \times 2$ lattices, which is inside the
deconfined phase.  We have done this for the full configurations,
and also for the vortex-only and vortex-removed configurations,
defined as in the zero-temperature case.  The results are shown in
Figs.\ \ref{hight-rho} and \ref{hight-f}.  The striking feature of
these figures is their strong resemblance to the corresponding
figures in the confined phase at zero temperature, namely Figs.\
\ref{r} and \ref{f} for full configurations, Figs.\ \ref{rcp} and
\ref{fcp} for vortex-only configurations, and Figs.\ \ref{rnv} and
\ref{fnv} for the vortex-removed configurations. Although we have
not attempted to determine the critical exponents in the deconfined
phase, it is clear that there is an enhanced density of low-lying
eigenvalues of the Faddeev--Popov operator, and that this is
associated with the center-vortex content of the configuration. It
is hard to avoid the conclusion that the Gribov scenario and the
vortex-dominance theory apply in both the confined and deconfined
phases.

\FIGURE[t]{
\centerline{\includegraphics[width=8truecm]{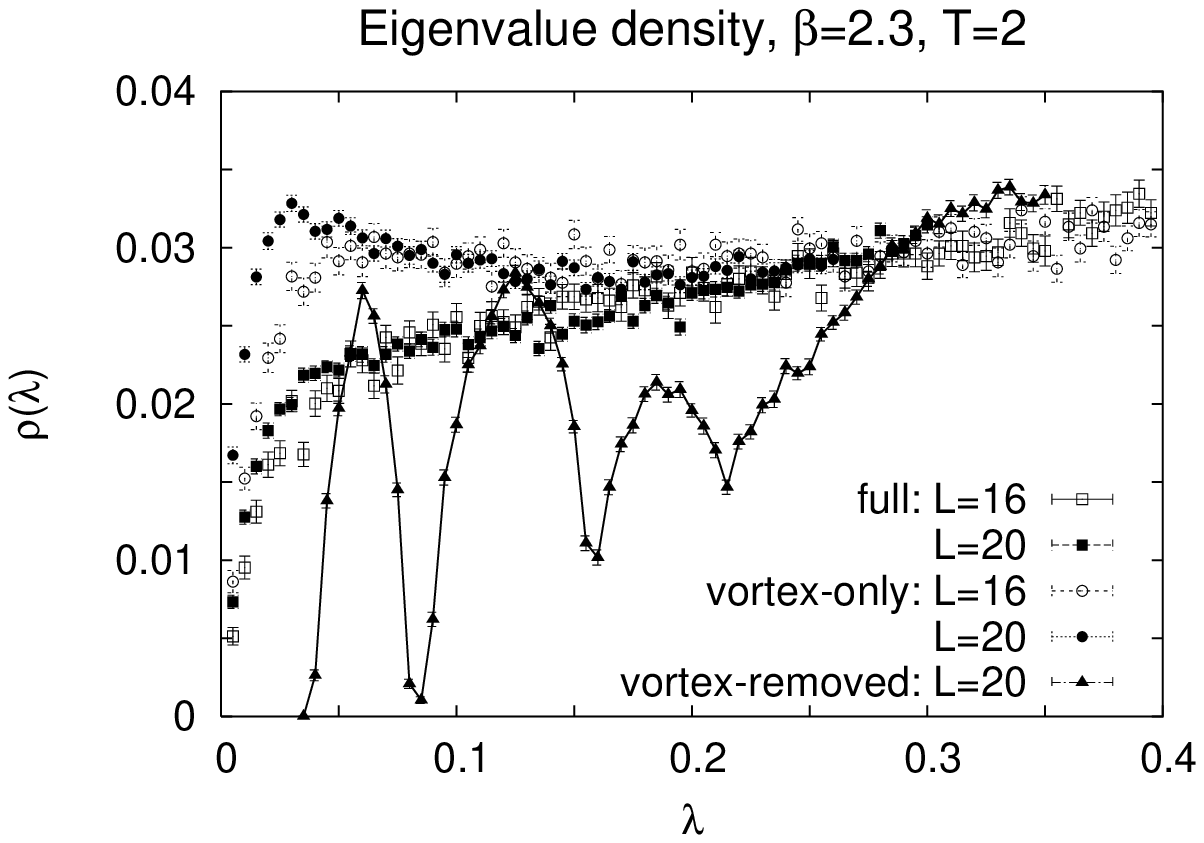}}
\caption{The F-P eigenvalue density in the deconfined phase for
full, vortex-only, and vortex-removed configurations.}
\label{hight-rho}}

\FIGURE[b]{
\centerline{\includegraphics[width=8truecm]{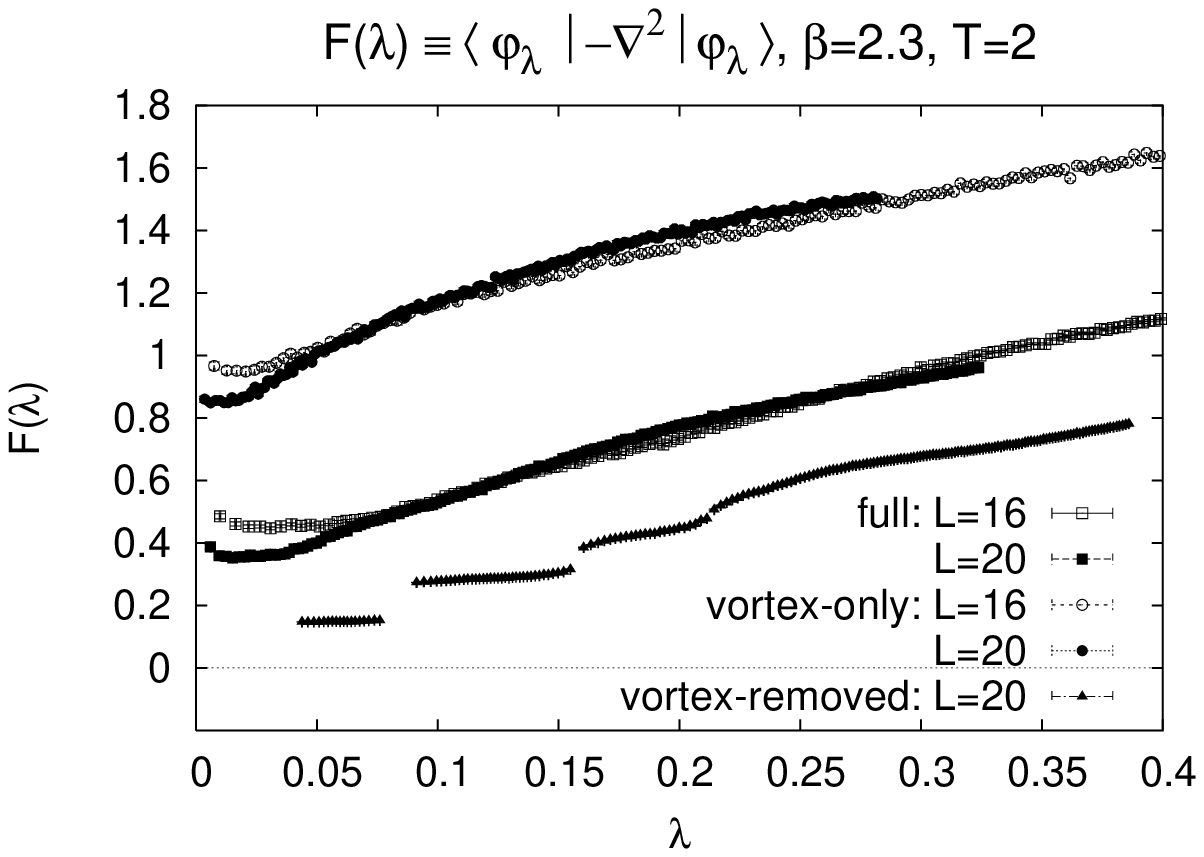}}
\caption{$F(\l)$ in the deconfined phase for full, vortex-only, and
vortex-removed configurations.} \label{hight-f}}

\section{Thin vortices and the eigenvalue density}\label{tvort}

     In the preceding sections we have studied the eigenvalue distribution in
thin vortex configurations, extracted via center projection from thermalized
lattices.  It is also of interest to study the eigenvalue spectrum of
thin vortex configurations of some definite geometry, such as an array.

    A single thin vortex, occupying two parallel planes in the four-dimensional
lattice, can be constructed in the following way:  Begin with the zero-field
configuration, i.e.\ all links equal to the identity matrix.  Then set
$U_2(x,y,z,t)=-I$ at sites
\begin{equation}
        1\le x \le {L\over 2},\quad y=1,\quad\mbox{all} ~ z,t.
\end{equation}
This creates two P-plaquettes in every $xy$ plane, which extends to
two vortex lines (stacks of P-plaquettes along a line on the dual
lattice) at a fixed time, and two vortex planes when extended also in
the $t$-direction of the lattice.  The two planes bound a connected
(Dirac) 3-volume, and will therefore be referred to as a single closed
vortex.

    Generalizing slightly, we create $N$ vortices parallel to the $zt$
plane by setting
\begin{equation}
        U_2(x,y_n,z,t) = -I
\end{equation}
for
\begin{equation}
     1\le x \le {L\over 2} ,
    \quad \left\{y_n = 1 + {n\over N}L,~n=0,1,\dots,N-1\right\}.
\end{equation}
In the same way (and making a few arbitrary choices of location), we create $N$
vortices parallel to the $xt$ plane by setting
\begin{equation}
        U_3(x,y,z_n,t) = -I
\end{equation}
for
\begin{equation}
    3\le y \le {L\over 2} + 2 ,
     \quad \left\{z_n = 1 + {n\over N}L,~n=0,1,\dots,N-1\right\}
\end{equation}
and parallel to $yt$ plane by
\begin{equation}
        U_1(x_n,y,z,t) = -I
\end{equation}
for
\begin{equation}
    {L\over 2}\le z \le L-1,
      \quad \left\{x_n = 1 + {n\over N}L,~n=0,1,\dots,N-1\right\}.
\end{equation}
We will consider a class of configurations characterized by the pair
of integers $(N,P)$, where $P$ is the number of planar orientations
(i.e.\ $P=1$ means only vortices parallel to the $zt$ plane, while
$P=3$ means vortices parallel to the $xt,~yt$ and $zt$ planes), and
$N$ is the number of closed vortices in each planar orientation.

\FIGURE[t]{
\centerline{\includegraphics[width=6truecm,angle=270]{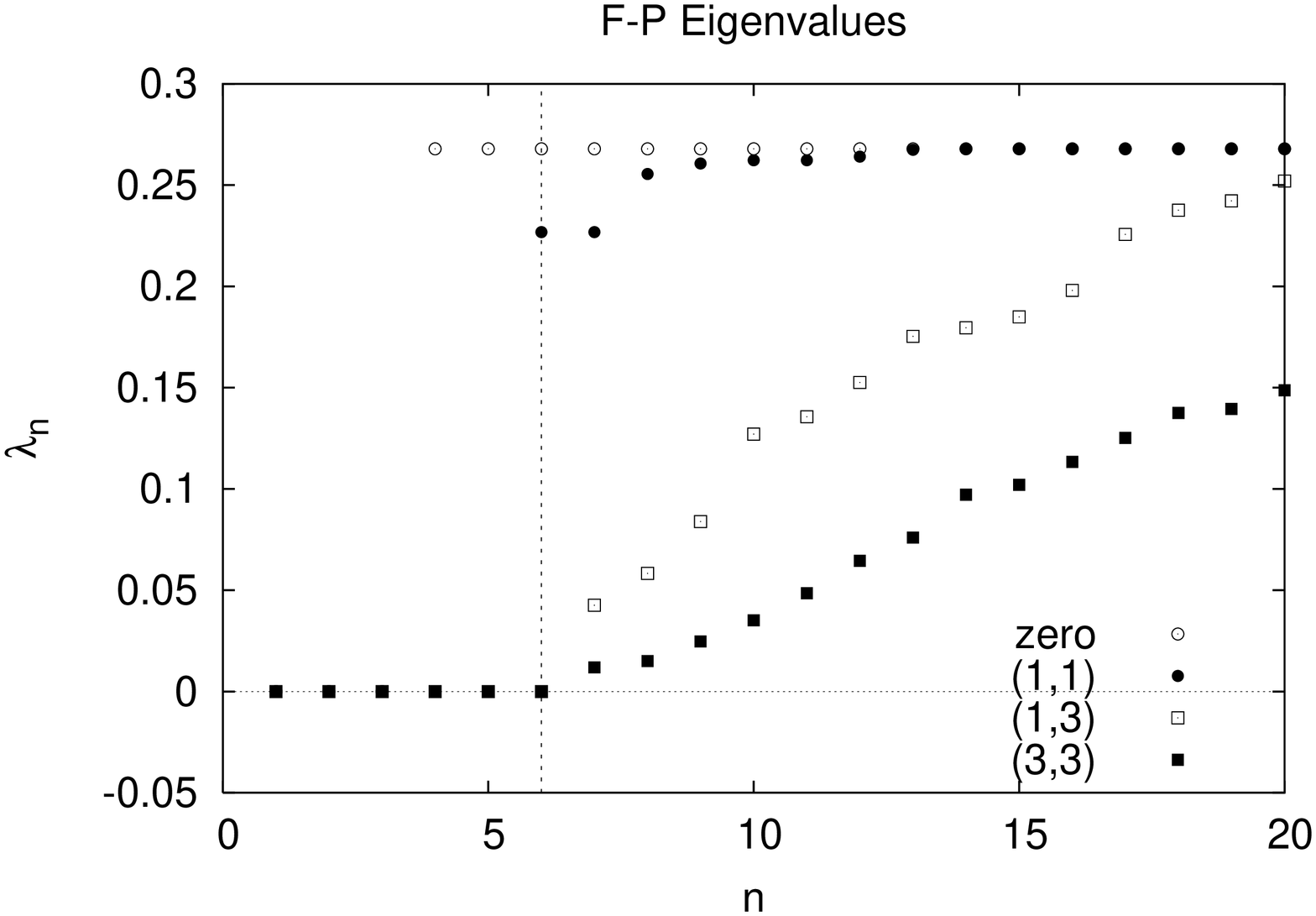}}
\caption{Distribution of low-lying eigenvalues in zero-field (0,0),
one-vortex (1,1), and vortex array (1,3), (3,3) configurations.}
\label{fp20}}

    As already noted, any configuration consisting of links equal to
$\pm I$ trivially fulfills the Coulomb gauge condition, but also
generally lies outside the Gribov region.  We therefore perform a
random gauge transformation on each thin vortex configuration, and
fix to a Coulomb gauge copy in the Gribov region by
over-relaxation.  Then the F-P eigenvalue spectrum (the same at any
time-slice) is extracted by the Arnoldi algorithm.

     The results are shown in Fig.\ \ref{fp20} for the first
20 F-P eigenvalues obtained on a $12^4$ lattice for zero-field
(0,0), one vortex (1,1), three vortex (1,3), and nine vortex (3,3)
configurations. As mentioned previously, it can be shown
analytically that arrays of thin vortices can have some additional
zero modes, beyond the three trivial zero modes which exist for
any lattice configuration.  These additional modes are seen
numerically in our calculation for the vortex spectrum.  Apart
from these extra zero modes, we see that the low-lying eigenvalue
spectrum of the one-vortex configuration is not much different
from the zero-field result.  But the magnitudes of the low-lying
eigenvalues changes drop abruptly, compared to the laplacian (0,0)
eigenvalues, upon going to a three vortex configuration, and drop
still further in a nine vortex array.  This is only a qualitative
result, but it does illustrate quite clearly the connection
between vortices and the Gribov horizon.  When the vortex geometry
is chosen to imitate various features of percolating vortices,
e.g.\ piercing planes in all directions and distributed throughout
the lattice, then the low-lying eigenvalues have very small
magnitudes as compared to the zero-field (or, in perturbation
theory, zero-th order) result. This implies a pileup of F-P
eigenvalues near $\l=0$, which we know is required for
confinement.

\section{Convexity of FMR and Gribov regions in SU(2) lattice gauge
theory}\label{convex}

In the section that follows this one, we will show that vortex
configurations play a special role in the geometry of the
fundamental modular region $\L$ and of the Gribov region $\Omega$.
But first we interrupt our narrative to establish a very general convexity
property of these regions in lattice gauge theory.

We start by recalling the well known convexity of $\L$ and $\Omega$ in
continuum gauge theory \cite{semenov}.  The Gribov region $\Omega$ is
the set of configurations that are transverse $\nabla \cdot A = 0$,
and for which the Faddeev--Popov operator $M(A) = - \nabla \cdot
\CD(A)$ is positive.  It is also the set of all relative minima on all
gauge orbits of the functional $F_A(g) = ||^{g}A||^2$, where $||A||$
is the Hilbert norm of the configuration~$A$, and $^{g}A$ is its gauge
transform by $g$.  The fundamental modular region $\L$ is the subset
of the Gribov region, $\L \subset \Omega$, which consists of all
absolute minima $F_A(1) \leq ||^{g}A||^2$ for all local gauge
transformations~$g(x)$.  It is well known that in continuum gauge
theory both of these regions are convex namely, if $A_1$ and $A_2$ are
configurations in $\L$ (or $\Omega$), then so is $A = \a A_1 + \b
A_2$, where $0 < \a < 1$, and $\b = 1 - \a$.  Geometrically, $A$ lies
on the line segment that joins $A_1$ and $A_2$.

We shall establish a similar, but slightly weaker property that holds
in SU(2) lattice gauge theory.  This is quite surprising because
convexity is a linear concept, whereas lattice configurations are
elements of a non-linear space.  We parametrize SU(2) configurations
by $U_i(x) = b_i(x) + i \vec{\sigma} \cdot \vec{a}_i(x)$, where
$b_i(x) = \pm [1 - \vec{a}^2_i(x)]^{1/2}$.  The $\vec{a}_i(x)$ would
be a complete set of coordinates, but for the sign ambiguity of
$b_i(x)$ above, corresponding to northern and southern hemisphere.  We
call ${\cal U}_+$ the set of configurations $U$ where all link
variables lie on the northern hemisphere
\begin{equation}
{\cal U}_+ \equiv \{ U:  b_i(x) = + [1 - \vec{a}^2_i(x)]^{1/2} \},
\end{equation}
for all $x$ and $i$.
The $\vec{a}_i(x)$ are a complete set of coordinates in  ${\cal U}_+$.

In minimal Coulomb gauge the fundamental modular region $\Lambda$ has
the defining property that a configuration $U$ in $\Lambda$ satisfies
\begin{equation}
\label{definefmr}
\sum_{xi} \mbox{Re Tr}\left[g(x)^{-1} U_i(x) g(x+\hat{\imath})\right]
\leq \sum_{xi} \mbox{Re Tr}\;U_i(x)
\end{equation}
for all $g(x)$.
Thus the gauge choice makes all the $U_i(x)$ as close to the identity
as possible, in an equitable way over the whole lattice.  In this
gauge, the link variables $U_i(x)$ for equilibrated configurations lie
overwhelmingly in the northern hemisphere, especially in the continuum
limit $\beta \rightarrow \infty$.  We define $\Lambda_{+}$ to be the
restriction of $\Lambda$ to ${\cal U}_+$, the set of configurations
whose link variables all lie in the northern hemisphere
\begin{equation}
\Lambda_+ \equiv \Lambda \cap {\cal U}_+,
\end{equation}
and we call it the restricted fundamental modular region.  We expect
that these are the important configurations in the continuum limit.
This is in fact necessary if the gauge-fixed lattice theory possesses
a continuum limit.  In this case $\vec{a}_i(x) \rightarrow -2 a
\vec{A}_i(x)$, where $a$ is the lattice spacing, and $\vec{A}_i(x)$ is
the continuum gauge connection.

         In the minimal lattice Coulomb gauge, the $\vec{a}_i(x)$
satisfy the lattice transversality condition
\begin{equation}
[{\rm div} \ a]^b(x) \equiv
\sum_i  \left[a_i^b(x) - a_i^b(x-\hat{\imath}) \right] = 0.
\end{equation}
This is a linear condition on the coordinates $a_i^b(x)$.
This suggests that we identify configurations in ${\cal U}_+$ with the
space of coordinates $a_i^b(x)$, and that we add configurations by
adding coordinates
\begin{equation}
\vec{a}_i(x) =
[\alpha \vec{a}_1 + \beta \vec{a}_2]_i(x) =  \alpha \vec{a}_{1,i}(x)
+ \beta \vec{a}_{2,i}(x).
\end{equation}
This yields a well-defined configuration $U \in {\cal U}_+$ only if
$\vec{a}_i^2(x) \leq 1$ for all links.  This is assured for the case
that is of interest to us where $0 < \alpha < 1$ and $\beta = 1 -
\alpha$.  Indeed, by the triangle inequality we have
\begin{equation}
|\vec{a}_{i}(x)| \leq |\alpha \vec{a}_{1,i}(x)| + |\beta \vec{a}_{2,i}(x)|
\leq \alpha + \beta = 1.
\end{equation}

{\it Statement:} In SU(2) lattice gauge theory the restricted
fundamental modular region is convex.  More precisely if $a_1$ and
$a_2$ lie in $\Lambda_+$, then $a \equiv \alpha a_1 + \beta a_2$ also
lies in $\Lambda_+$ for $0 < \alpha < 1$ and $\beta = 1 - \alpha$.

The proof is given in Appendix \ref{B}.

         We may establish the same convexity property for the Gribov
region $\Omega$, namely that $\Omega_+ \equiv \Omega \cap {\cal U_+}$
is convex.  The Gribov region has the defining property
\begin{eqnarray}
&&\sum_{xi} \ \left\{ b_{i}(x)
\left[\vec{\omega}_i(x+\hat{\imath}) - \vec{\omega}_{i}(x)\right]^2\right.
\\
\nonumber
&&\left.- \left[\vec{\omega}_i(x+\hat{\imath}) - \vec{\omega}_i(x)\right] \cdot
\vec{a}_i(x) \times
\left[\vec{\omega}_i(x+\hat{\imath}) +
\vec{\omega}_i(x)\right]\right\}  \geq 0,
\end{eqnarray}
for all $\omega$.  The proof for $\Omega$ is the same as for $\L$
because this inequality has the same structure as \rf{definefmr},
being linear in $a_i(x)$ and $b_i(x)$.

\section{Vortices as vertices}\label{vv}

We have seen that $\Omega_+$ is convex, so one might think it has a
simple oval shape.  We shall show however that when thin center vortex
configurations are gauge transformed into minimal Coulomb gauge they
are mapped into points $U_0$ on the boundary $\partial \Omega$ of the
Gribov region $\Omega$ where this boundary has a wedge-conical
singularity of a type that is described below.  This illustrates the
intimate relation of dominance by configurations in the neighborhood
of the Gribov horizon or in the neighborhood of thin center vortex
configurations.

\subsection{Previous results}

We call a ``thin center vortex configuration" (or center
configuration) one where, in maximal center gauge, all link variables
are center elements $U_i(x) = Z_i(x)$.  [Such a configuration may be
characterized in a gauge-invariant way by the statement that all
holonomies are center elements.  (A holonomy is a closed Wilson loop.)
This definition also holds in continuum gauge theory.]
%where the maximal center gauge does not exist.]
When any configuration is gauge transformed by a minimization
procedure, such as used here, it is mapped into the Gribov region
$\Omega$.  In \cite{Us} it was shown: {\it When a thin center vortex
configuration is gauge transformed into minimal Coulomb gauge it is
mapped onto a configuration $U_0$ that lies on the boundary $\partial
\Omega$ of $\Omega$.  Moreover its Faddeev--Popov operator $M(U_0) = -
\nabla \cdot \CD(U_0)$ has a non-trivial null space that is $(N^2
-1)$-dimensional.}  (This is in addition to the $(N^2 -1)$-dimensional
trivial null-space consisting of constant eigenvectors.  Here and
below we do not count trivial eigenvectors that are generators of
global gauge transformations.)  Likewise, when an abelian
configuration is gauge-transformed into the Gribov region, its
Faddeev--Popov operator has a non-trivial $R$-dimensional null space,
where $R$ is the rank of the group.  The reason is that the gauge
orbit of a center or abelian configuration is degenerate so the
gauge-invariant equation $\CD_i(U_0) \omega = 0$, which holds for $i =
1,2,3$ simultaneously, has $N^2-1$ or $R$ non-trivial solutions
respectively.

\subsection{Tangent plane to Gribov horizon at a regular point}

Let $a_i^b(x)$ be a set of coordinates of the group element $U_i(x) =
U[a_i^b(x)]$.  It is convenient to use coordinates $a_i^b(x)$ such
that the Coulomb gauge condition is linear, $(\nabla \cdot a)^b(x) =
\sum_i\;[a_i^b(x)-a_i^b(x-\hat{\imath})] = 0$, and in terms of which
$\Omega_+$ (see above) is convex.  We write $U_0 + \delta U = U(a_0 +
\delta a)$.  Here $\delta a$ is an arbitrary (transverse) small
variation of the coordinates at $a_0$.  Such a variation is a tangent
vector at $a_0$, and the space of tangent vectors constitutes the
tangent space at $a_0$.

Let $U_0$ be a configuration in Coulomb gauge, that lies on the
boundary $\partial \Omega$ of the Gribov region.  By definition, the
corresponding Faddeev--Popov operator has a non-trivial null
eigenvector
\begin{equation}
M(U_0) \omega_0 = 0,
\end{equation}
all other eigenvalues being non-negative.  We are interested in the
case of a center configuration in Coulomb gauge where the non-trivial
null eigenvalue of $M(U_0)$ is $(N^2-1)$-fold degenerate, but for
orientation we first consider the case where the non-trivial null
eigenvalue is non-degenerate.  Let $U_0 + \delta U$ be a neighboring
point that is also on the Gribov horizon, so $M(U_0 + \delta U)$ also
possesses a null vector
\begin{equation}
M(U_0 + \delta U) (\omega_0 + \delta \omega) = 0.
\end{equation}

We wish to find the condition on $\delta a$ that holds whenever $U(a_0
+ \delta a)$ also lies on the Gribov horizon $\partial\Omega$.  We
have
\begin{equation}
M(U_0 + \delta U) = M(U_0) + \delta M,
\end{equation}
where
\begin{equation}
\delta M = \sum_{xib}\delta a_i^b(x) \ \partial_{xib} M_0 ,
\end{equation}
and
\begin{equation}
\partial_{xib} M_0 \equiv \left.{ { \partial M } \over { \partial
a_i^b(x) } }\right\vert_{a = a_0}.
\end{equation}
To first order in small quantities we have
\begin{equation}
  \delta M \ \omega_0 + \ M(U_0) \delta \omega = 0.
\end{equation}
We contract this equation with $\omega_0$ and obtain
\begin{equation}
  (\omega_0, \delta M \ \omega_0) = 0,
\end{equation}
or
\begin{equation}
\sum_{xib} \delta a_i^b(x) \ (\omega_0, \partial_{xib}M_0 \ \omega_0)
  = 0.
\end{equation}
Geometrically, this is the statement that $\delta a_i^b(x)$ is
perpendicular to the vector $(\omega_0, \partial_{xib}M_0 \
\omega_0)$, so this vector is the normal to $\partial\Omega$ at $U_0 =
U(a_0)$.  It defines the a hyperplane that is tangent to the Gribov
horizon at $a_0$.

\subsection{Center vortices as singularities of the Gribov horizon}

\subsubsection{General idea}

Suppose that $U_0 = U(a_0)$ is a point on the Gribov horizon and that
the null eigenvalue of $M(a_0)$ is $P$-fold degenerate,
\begin{equation}
\label{nulleigen}
M(a_0) \ \omega_0^{(n)} = 0; \qquad n = 1,\dots, P.
\end{equation}
As noted above, this happens for every gauge copy in Coulomb gauge of
a thin center vortex configuration, where $P = N^2 -1$.  Under the
small perturbation $M(a_0 + \delta a) = M(a_0) + \delta M$, where
$\delta M$ is given above, the $P$-fold degenerate null-space splits
into $P$ levels with eigenvalues $\delta\lambda_n$, for $n = 1,\dots,
P$, that depend linearly, $\delta\lambda_n = \sum_{xia} C_{n,i}^b(x)
\delta a_i^b(x)$, on the $\delta a$.

For a point $a_0$ on the boundary $\partial\Omega$ we define the
Gribov region of the tangent space at $a_0$ to be the set of tangent
vectors that point inside $\Omega$.  We designate it by
$\Omega_{a_0}$.  More formally, it is the set of tangent vectors $\d
a$ at $a_0$ such that $a_0 + \delta a$ lies in~$\Omega$,
\begin{equation}
\Omega_{a_0} \equiv \{\delta a: a_0 + \delta a \in \Omega \}.
\end{equation}
Recall that by definition the Gribov region $\Omega$ consists of
(transverse) configurations $U(a)$ such that all eigenvalues of $M(a)$
are positive.  It follows that $\Omega_{a_0}$ is the set of $\delta a$
such that all $P$ eigenvalues $\delta\lambda_n(\delta a)$ are
positive,
\begin{equation}
\Omega_{a_0} \equiv \{ \delta a: \delta \lambda_n(\delta a) > 0 \ \
{\rm for} \ \ n= 1,\dots , P\}.
\end{equation}
This condition is quite restrictive because, for generic $\d a$, some
number $\nu$ of eigenvalues $\d \l_n$ are negative while $P - \nu$ are
positive, where $\nu = 0, 1, \dots , P$.  As a result the boundary
$\partial \Omega_{a_0}$ of the Gribov region at $a_0$ is not simply a
tangent plane through~$a_0$, as before.  Rather it is a high
dimensional wedge-conical vertex whose shape we shall find.

\subsubsection{Degenerate perturbation theory}

         The eigenvalues in an infinitesimal neighborhood of a point
$U_0$ on the Gribov horizon are determined by the eigenvalue equation
\begin{equation}
  [ \ M(U_0) + \delta M ] \left( \ \sum_n c_n \omega_0^{(n)} + \delta
  \omega \right) = \d \l \left( \ \sum_n c_n \omega_0^{(n)} + \delta
  \omega \ \right),
\end{equation}
or
\begin{equation}
  \delta M \sum_n c_n \omega_0^{(n)} \ + \ M(U_0) \delta \omega \ = \d
\l \ \sum_n c_n \omega_0^{(n)}.
\end{equation}
Upon contracting this equation with $\omega_0^{(m)}$, we obtain, by
\rf{nulleigen}, the $P$-dimensional eigenvalue equation
\begin{equation}
\sum_n \delta a_{mn} \; c_n = \d \l \ c_m,
\end{equation}
where
\begin{eqnarray}
\delta a_{mn} & \equiv & \left(\omega_0^{(m)}, \delta M \
\omega_0^{(n)}\right) \cr & = & \sum_{xib} \delta a_i^b(x) \
\left(\omega_0^{(m)}, \partial_{xib}M_0 \ \omega_0^{(n)}\right).
\end{eqnarray}
Abstractly, $\d a$ is a tangent vector in lattice configuration space
at point $a_0$ in lattice configuration space; the components of this
vector, in a suitable basis, are denoted $\d a^b_i(x)$.  We may also
regard the set of numbers $\d a_{mn}$ as components of $\d a$ in some
other basis.

The eigenvalue equation has $P$ solutions $\d \l_n$.  The condition
that they all be positive, which determines $\Omega_{a_0}$, the Gribov
region at $a_0$, is the condition that the matrix $\delta a_{mn}$
define a strictly positive form.  Such a form is characterized by the
Sylvester criterion that $\det \delta a_{mn}$ be positive, together
with the determinant of all principle minors (diagonal square
submatrices) of this matrix.  The boundary of this region is
determined by the condition that one eigenvalue vanish, which happens
when the determinant vanishes,
\begin{equation}
\det \delta a_{mn} = 0.
\end{equation}

\subsubsection{Two-fold degeneracy}

We first analyze 2-fold degeneracy.  In this case positivity of the
two principle minors and of the determinant reads
\begin{equation}
\d a_{11} > 0, \qquad \d a_{22} > 0, \qquad \delta a_{11} \ \delta
a_{22} - \delta a_{12}^2 > 0.
\end{equation}
In terms of $\delta a_+ \equiv\frac{1}{2} (\delta a_{11} + \delta
a_{22})$ and $\delta a_- \equiv\frac{1}{2} (\delta a_{11} - \delta
a_{22})$, the last condition reads
\begin{equation}
\delta a_+^2 - \delta a_-^2 - \delta a_{12}^2 > 0.
\end{equation}
The three inequalities define the interior of the ``future" cone in
the 3-variables $\delta a_+$, $\delta a_-$ and $\delta a_{12}$, with
vertex at the origin.  Thus the boundary of the Gribov region at $a_0$
is a cone in these 3 variables.  Taking account of the remaining
components of $\d a$, the conical singularity can be viewed as a
kind of wedge in higher dimensions.

\subsubsection{Three-fold degeneracy}

For SU(2) gauge theory the thin-vortex configurations are 3-fold
degenerate points on the Gribov horizon.  In this case the positivity
of the principle minors and the determinants reads
\begin{equation}
\delta a_{11} > 0, \qquad \delta a_{22} > 0, \qquad \delta a_{33} > 0
\end{equation}
\begin{eqnarray}
\delta a_{11} \ \delta a_{22} - \delta a_{12}^2 & > & 0 \cr \delta
a_{11} \ \delta a_{33} - \delta a_{13}^2 & > & 0 \cr \delta a_{22} \
\delta a_{33} - \delta a_{23}^2 & > & 0
\end{eqnarray}
\begin{equation}
\det \delta a_{mn} > 0.
\end{equation}
These 7 inequalities characterize the Gribov region $\Omega_{a_0}$ of
the tangent space at a point $a_0$ on the Gribov horizon that is
3-fold degenerate.

From our discussion of 2-fold degeneracy we know that the positivity
of each of the 2 by 2 determinants and of its diagonal elements
defines a ``future" cone.  We call these future cones $F_{12}$,
$F_{13}$ and $F_{23}$ respectively.  For example $F_{12}$ is defined
by $\delta a_{11} > 0$, and $\delta a_{22} > 0$, and
\begin{equation}
\left[{\textstyle{\frac{1}{2}}}(\delta a_{11} + \delta
a_{22})\right]^2 - \left[{\textstyle{\frac{1}{2}}}(\delta a_{11} -
\delta a_{22})\right]^2 - \delta a_{12}^2 > 0,
\end{equation}
etc.  Each is a cone of opening half-angle $\pi/4$, and we have shown
that $\Omega_{a_0}$ is contained in the intersection of the 3 future
cones,
\begin{equation}
\Omega_{a_0} \subset F_{12} \cap F_{13} \cap F_{23}.
\end{equation}
This condition and the 3 by 3 determinantal inequality $\det \delta
a_{mn} > 0$ characterize the Gribov region $\Omega_{a_0}$ in the
tangent space of a point $U_0 = U(a_0)$ that is 3-fold degenerate.

\subsubsection{Over-all picture of the Gribov horizon and its
center-vortex singularities}

We have seen that $\Omega_+$ is convex and that thin center vortex
configurations are wedge-conical singularities on the boundary of
$\partial \Omega$.  Those that are on $\partial \Omega_+$ are extremal
elements, like the tips on a very high-dimensional pineapple.  Indeed,
each center configuration is an isolated point.  If one moves a small
distance from a center configuration it is no longer a center
configuration.  Its gauge transform $U(a_0)$ in Coulomb gauge is
likewise an isolated point on the Gribov horizon.  Thus the wedge in
the boundary $\partial\Omega$ at $a_0$ that we have just described
occurs at an isolated point where the Gribov horizon may be said to
have a ``pinch".

In SU(2) gauge theory there are $2^{d V}$ center configurations (where
$d$ is the number of dimensions of space, and $V$ is the volume of the
lattice) because there are $d V$ links in the lattice and there are 2
center elements in SU(2).  These are related by $2^V$ gauge
transformations, so there are $2^{(d-1)V}$ center orbits.  The
absolute minimum of each of these orbits lies on the common boundary
of the fundamental modular region $\partial \Lambda$ and the Gribov
region $\partial \Omega$.  So there are at least $2^{(d-1)V}$ tips on
the above-mentioned pineapple, a truly enormous number.  Moreover for
each such orbit there are many Gribov copies, all lying on $\partial
\Omega$ (spin glass problem).  These are all singular points of the
Gribov horizon of the type described.  For SU(2) there may not be any
other singular points on $\partial\Omega$.  It is possible that the
thin center vortex configurations provide a rather fine triangulation
of~$\partial \Omega$.

Note: The gauge transform in Coulomb gauge of an abelian configuration
also lies on the Gribov horizon.  However the SU(2) group is of rank
1, so, for SU(2) gauge theory, such a configuration is invariant under
a one-parameter group of transformations only.  As a result, the
corresponding null space $M(U_0)\omega_0 = 0$ is only one-dimensional,
and the present considerations do not indicate that these are
singularities of~$\partial\Omega$.

\section{Coulomb gauge as an attractive fixed point of the
interpolating gauge}\label{cgfp}

    The data reported above have been obtained by numerically gauge
fixing to the minimal Coulomb gauge using a well-defined numerical
minimization procedure on a lattice of finite volume.  However not
every lattice gauge corresponds to a well-defined continuum gauge, so
one may well ask whether our results have a continuum analog,
especially since in continuum theory the Coulomb gauge is a singular
gauge.  Indeed, in Coulomb gauge, some propagators of the form $1/{\bf
k}^2$ appear instead of $1/(k_0^2 + {\bf k}^2)$, which leads to
unrenormalizable divergences $\int dk_0$ in some closed loops.  This
has been treated by using the action in phase space (first order
formalism), which allows a systematic cancellation of these
divergences \cite{renormcoul}, or by using an interpolating gauge
\cite{BZ} with gauge condition
\begin{equation}
\label{interpolate}
a \partial_0 A_0 + \nabla \cdot {\bf A} = 0.
\end{equation}
This gauge interpolates between the Landau gauge $a = 1$ and the
Coulomb gauge, $a = 0$, and may also be achieved by a numerical
minimization procedure.  For $a > 0$ it is a regular continuum gauge,
and the gauge parameter $a$ provides a regularization of the
divergences of the Coulomb gauge, which is obtained in the limit $a
\rightarrow 0$ of the interpolating gauge.

However as a possible obstacle to this regularization, it must be
noted that the quantities that appear in the interpolating gauge
condition \rf{interpolate} are unrenormalized quantities, and the
unrenormalized gauge parameter depends on the ultraviolet
cut-off~$\L$,
\begin{equation}
a = a(\Lambda/\Lambda_{\rm QCD}).
\end{equation}
For the Coulomb-gauge limit of the interpolating gauge to be a smooth
limit it is necessary that the gauge parameter $a(\Lambda/\Lambda_{\rm
QCD})$ flows {\it toward} (or at least not away from) the Coulomb-gauge
value $a = 0$, as the ultraviolet cut-off is removed, \mbox{$\L
\rightarrow \infty$}.  In Appendix \ref{C} its dependence on $\L$ is
calculated in the neighborhood of $a = 0$ to one-loop order in the
perturbative renormalization group, with the result in pure SU($N$)
gauge theory,
\begin{equation}
a(\Lambda/\Lambda_{\rm QCD}) = { { {\rm const} } \over {
[\ln(\Lambda/\L_{\rm QCD})]^{4/11} } },
\end{equation}
This gives
\begin{equation}
\lim_{\Lambda \rightarrow \infty} a(\Lambda/\Lambda_{\rm QCD}) = 0,
\end{equation}
and we conclude that in some neighborhood of $a = 0$, the Coulomb
gauge value $a = 0$ is an attractive fixed point of the
renormalization-group flow. This removes the possible obstacle to
regularization of the Coulomb gauge by the interpolating gauge.

[Similarly the renormalized gauge parameter $a^R =
a^R(\mu/\Lambda_{\rm QCD})$, depends on the renormalization
mass~$\mu$.  As the renormalization mass $\mu$ gets large, which is
the appropriate choice for calculations at high momentum, one has at
one-loop order
\begin{equation}
a^R(\m/\Lambda_{\rm QCD}) = { { {\rm const} } \over { [\ln(\m/\L_{\rm
QCD})]^{4/11} } }.
\end{equation}
So again, the Coulomb-gauge value $a^R = 0$ is an attractive fixed
point of the renormalization-group flow.]

\section{Conclusions}

    We have found that the low-lying eigenvalues of the Faddeev--Popov
operator, in thermalized lattices, tend towards zero as the lattice
volume increases.  This means that in the infinite volume limit,
thermalized configurations lie on the Gribov horizon.  That fact alone
would not allow us to make any strong conclusions about the energy of
unscreened color charge.  However, the data also indicate that the
\emph{density} $\r(\l)$ of F-P eigenvalues goes as a small power of
$\l$, at infinite volume, as $\l\ra 0$.  Together with the behavior
of $F(\l)$
at $\l\ra 0$, we conclude that the energy of an unscreened color
charge is infrared divergent, and that this divergence can be
attributed to the near-zero modes of the Faddeev--Popov operator.

    This evidence clearly supports the Gribov horizon scenario
advocated by Gribov and Zwanziger in ref.\ \cite{horizon}.  This
scenario was invented to account for confinement, and the reader may
be surprised to find a linearly rising color-Coulomb potential in the
{\it deconfined} phase.
However this is nicely explained by the horizon scenario in Coulomb
gauge, as we now explain.  In Coulomb gauge the gauge fixing is done
independently on each 3-dimensional time slice, and may be done on a
single time slice.  According to the horizon scenario, on each time
slice, 3-dimensional configurations $A_i({\bf x})$ are favored that
lie near the horizon of a 3-dimensional gauge theory, and this
enhances the instantaneous color-Coulomb potential.  This is true for
{\it every} temperature $T$, including in the deconfined phase,
because temperature determines the extent of the lattice in the {\it
fourth} dimension.  Thus, the
horizon scenario provides a framework in which confinement may be
understood, but it is not detailed enough to tell us under what
conditions the infinite color-Coulomb potential may be {\it screened}
to give a finite self-energy, as measured by the Polyakov loop.

    By factoring thermalized lattices into vortex-only and
vortex-removed components, we have also been able to show that the
constant density of low-lying F-P eigenvalues can be entirely
attributed to the vortex component.  We find that the eigenvalue
density of the vortex component is qualitatively similar to that of
the full configuration. The eigenvalue density of the vortex-removed
component, on the other hand, is drastically different from that of
the full configuration.  This density can be interpreted as simply a
small perturbation of the zero-field (or zero-th order) result, and it
is identical in form to the (non-confining) eigenvalue density of
lattice configurations in the Higgs phase of a gauge-Higgs theory.

    These findings establish a firm connection between the center
vortex and the Gribov horizon confinement scenarios.  According to the
center vortex doctrine, fluctuations in the vortex linking number are
responsible for the area law falloff of Wilson loops. It now appears
that vortex configurations are also responsible for the enhanced
density of near-zero F-P eigenvalues, which is essential to the Gribov
horizon picture.  This result is consistent with previous results in
ref.\ \cite{JS}, where it was found that vortex removal also removes
the Coulomb string tension of the color Coulomb potential.
It is also consistent with recent investigations of Gattnar, Langfeld,
and Reinhardt \cite{Gattnar} in Landau gauge.

    The F-P eigenvalue spectrum at high temperatures, with and without
vortices, turns out quite similar to the corresponding results at
low-temperature. This similarity was to be expected, since the F-P
operator depends only on spacelike links at a fixed time, and even
at high $T$ these form a confining three-dimensional ensemble for
spacelike Wilson loops. The Gribov scenario must therefore be
relevant to physics in the deconfined phase; cf.\ ref.\ \cite{Dan2}
for a recent application.

We also report a result which supports the consistency of Coulomb
gauge itself in the continuum limit.  Coulomb gauge is very singular
in continuum perturbation theory, and one method of making it better
defined is to view Coulomb gauge as a non-singular limit of the more general
gauge condition $a \pa_0 A_0 + \nabla \cdot {\bf A} = 0$.  The success
of this approach depends on whether the Coulomb gauge limit, $a=0$, is
an attractive ultraviolet fixed point of the renormalization group
flow.  Here we have shown that this requirement is satisfied.

    Finally, we have uncovered an intriguing geometrical property of thin
vortices in lattice configuration space.  In ref.~\cite{Us} it was
shown that thin vortices (gauge equivalent to center configurations)
lie on the Gribov horizon.  The Gribov horizon is a convex manifold,
and we have shown here that thin vortices are conical singularities
on that manifold.  Percolating thick vortices appear to be
ubiquitous in thermalized lattice configurations at or near the
Gribov horizon; it is conceivable that the special geometrical
status of thin vortices is in some way related to the ubiquity of
their finite-thickness cousins.

%
% Acknowledgements
%
\acknowledgments{%
   J.G.\ thanks Poul Henrik Damgaard for helpful discussions
on the scaling of matrix model eigenvalue distributions.

Our research is supported in part by the U.S. Department of Energy
under Grant No.\ DE-FG03-92ER40711 (J.G.), the Slovak Science and
Technology Assistance Agency, Grant No.\ APVT--51--005704
(\v{S}.O.), and the National Science Foundation, Grant No.\
PHY-0099393 (D.Z.). }

\appendix

\section{Finite-volume scaling of low-lying F-P eigenvalues}
\label{A}

    In $N\times N$ random matrix models, the tail of the eigenvalue
distribution often displays a universal scaling behavior with $N$.
This fact has found important applications in the study of chiral
symmetry breaking, where this sort of universal scaling is found in
the density of near-zero eigenvalues of the Euclidean Dirac operator
as a function of lattice volume (which is proportional to the number
of eigenvalues) \cite{Verb}.  In our case, we are also interested in
the density of near-zero eigenvalues in the infinite 3-volume limit.
The statement in this case is as follows (cf., e.g., ref.\
\cite{Janik}): Suppose the normalized density of low-lying
eigenvalues, at very large volumes, goes as
\begin{equation}
         \r(\l) = \k \l^\a
\end{equation}
where $\k$ is some constant.
Then the density of eigenvalues, the average spacing between the
low-lying eigenvalues, and the
probability density $P(\l_n)$ of the $n$-th low-lying eigenvalue, agree
for every lattice 3-volume $V_3$, if the eigenvalues
themselves are rescaled according to
\begin{equation}
         z = \l V_3^{1\over 1+\a} .
\label{rescale}
\end{equation}
The argument goes as follows:  The number of eigenvalues $N[\l,\D \l]$
in the interval $[\l-\oh \D \l,\l+\oh \D \l]$ is
\begin{equation}
       N[\l,\D \l] = 3V_3 \r(\l) \D \l .
\end{equation}
Then in terms of rescaled eigenvalues $z=\l V_3^p$, we have
\begin{equation}
        N[\l,\D \l] = 3\k V^{1-p(1+\a)}_3 z^\a \D z .
\end{equation}
If we require that this number of eigenvalues depends only on
the rescaled variables $z,~\D z$, then it is necessary that
$p=1/(1+\a)$, and eq.\ \rf{rescale} follows.

   Our strategy is to plot the probability density $P(\l_{min})$,
rescaled by a factor $V_3^{-1/(1+\a)}$, as a function of the variable
$z_{min} = \l_{min} V_3^{1/(1+\a)}$, where $\l_{min}$ is the lowest
non-zero eigenvalue.  The rescaling of the probability density ensures
that its integral over $z_{min}$ is unity.  $P(\l_{min})$ is computed
on $8^4$ to $20^4$ lattice volumes, at various values of $\a$.  If we
can find a value of $\a$ for which the (rescaled) probability
densities $P(\l_{min})$ coincide at all lattice sizes, then this
determines $\a$, and in turn the behavior of the eigenvalue density
near $\l=0$.

    The results for rescaled $P(\l_{min})$, for both the full configurations
($\l_{min}=\l_4$), and the vortex-only  configurations ($\l_{min}=\l_7$),
are displayed in Fig.\ \ref{rho_le}.  We show only three
values of $\a$ in each case, which include our best estimate for the scaling
$\a$. We find that $\a=0.25 \pm 0.05$
for the full configurations, and $\a=0.0\pm 0.05$
for the vortex-only configurations seem to give the best scaling results;
the error estimate is subjective.

\FIGURE[h!]{
\centerline{\includegraphics[width=13.8truecm]{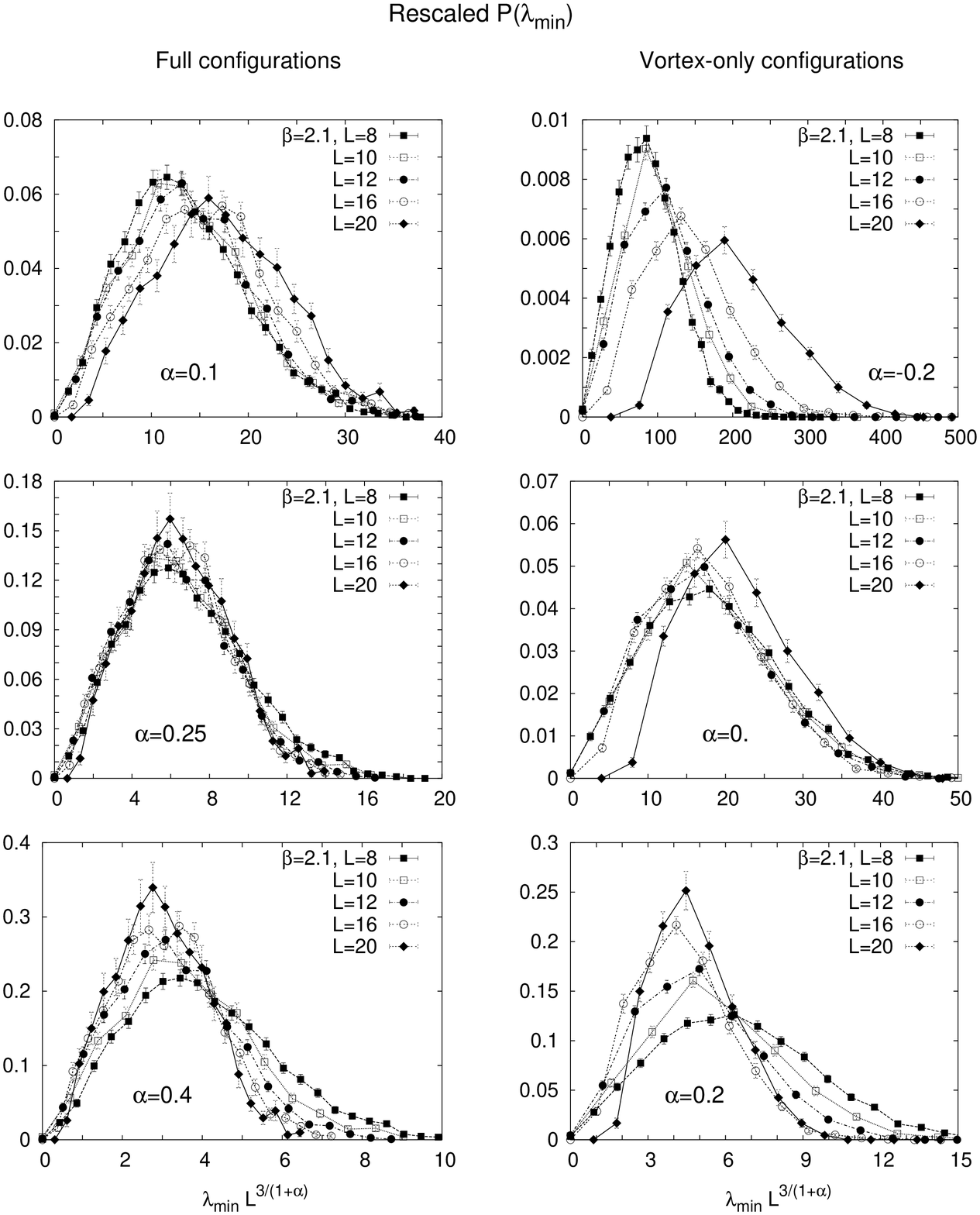}}
\caption{Probability distribution of the values of the lowest
non-trivial FP eigenvalue $\l_{min}$ on a variety of lattice sizes
$8^4-20^4$, as a function of the rescaled eigenvalue
$z_{min}=\l_{min} L^{3/1+\a}$.  Results for the full configurations
are in the left-hand column of figures, with the vortex-only results
in the right-hand column. Each figure corresponds to a different
choice of $\a$; $\a=0.1,0.25.0.4$ for the full configurations,
$\a=-0.2,0.0,0.2$ for the vortex-only configurations. The closest
match of rescaled $P(\l_{min})$ at different volumes comes at
$\a\approx 0.25$ for the full configurations, and $\a\approx 0$ for
vortex only configurations (with the exception of $L=20$, see
text).} \label{rho_le}}

     The lattice data for $L=20$ in the vortex-only configurations
calls for some comment.  This distribution does not match the
distributions obtained at smaller $L$ at any $\a$.
This could mean that the scaling
hypothesis for low-lying eigenvalues is invalid, but in our opinion
the mismatch at $L=20$ has a different explanation: We
believe the problem is connected to difficulties we have encountered
with gauge fixing on these large lattices.  As lattice size is increased,
typical values of $\l_{min}$ become smaller, and the number of over-relaxation
steps required to satisfy our Coulomb gauge convergence criterion increases.
In general, on a given lattice size, the number of over-relaxation steps
required to gauge fix is inversely correlated with the size of $\l_{min}$
in the gauge-fixed configuration.

    It happens occasionally that even after 10,000 gauge-fixing steps
on a given time-slice, the spatial lattice configuration has not
converged to Coulomb gauge.  When this happens, we perform a random
gauge transformation at that time slice and gauge fix a second time,
to a different Gribov copy (note that in Coulomb gauge, time-slices
can be, and were, gauge-fixed independently).  This procedure most
likely biases
the result towards higher average values of $\l_{min}$.  In doing
things this way, we are almost certainly modifying the probability
distribution in Gribov region, giving lower measure to Gribov copies
which are closer to the horizon.  On the other hand, simply excluding
these hard-to-gauge-fix lattices would probably introduce an even
worse bias. On smaller ($L<20$) vortex-only lattices, and on the full,
unmodified lattices, problems with convergence to Coulomb gauge are
uncommon.  However, on the $L=20$ vortex-only lattice, there is a
convergence failure on almost 38\% of all time slices; on these slices
we have performed a random gauge transformation to move to a different
Gribov copy.  The rate of convergence failure on vortex-only $16^4$
lattices is five times lower, and the rate of failure on unmodified
$20^4$ lattices is eight times lower, than on the vortex-only $20^4$
lattice.  For this reason, we believe that the bias
towards larger eigenvalues is by far the worst on the vortex-only
$20^4$ lattices, and this is in fact where we see the mismatch, in
$P(\l_{min})$ vs.\ $z_{min}$ at $\a\approx 0$, with the other vortex-only
lattice volumes.

    Note that for the vortex-only configurations, $\a=0$ is consistent
with a finite, non-zero value for $\rho(0)$, as shown in
eq.\ \rf{vo}.  On the other hand, it is not excluded that $\a$ could
in fact be slightly negative, in which case $\rho(\l)$ actually
diverges at $\l=0$.  Since the eigenvalue density in Fig.\ \ref{rcp}
does appear to actually rise as $\l \ra 0$, before suddenly falling, a
divergent behavior at $\l=0$ in the infinite volume limit is not at
all excluded.

    Next we consider $F(\l)$ as $\l\ra 0$.  We have fit the average
value of $F$ at the lowest non-zero eigenvalue,
$\langle F(\l_{min})\rangle$, to the form
\begin{equation}
       \langle F(\l_{min}) \rangle = {a \over L^p} + b .
\end{equation}
Our result, within errorbars, is consistent with
\begin{eqnarray}
       \langle F(\l_4)\rangle
            &=& {10\over L}   ~~~~~~~~~~~~ \mbox{full configurations,}
\non \\
       \langle F(\l_7) \rangle &=& 1 +{ 20\over L^{1.4}}
                ~~~~ \stackrel{\mbox{vortex-only}}
                              {\mbox{configurations.}}
\end{eqnarray}
Since $\l_{min}\ra 0$ at infinite volume,
then $F(0)=1$ for vortex-only configurations, as stated in eq.\ \rf{vo}.
For the full configurations, since we estimate $\a\approx 0.25$,
with the consequence that  $\langle \l_{min} \rangle \sim 1/L^{2.4}$,
it is reasonable to guess that for the full configurations near $\l=0$,
\begin{equation}
         F(\l) \sim \l^{0.42} .
\end{equation}
We have therefore tried the following fit
\begin{equation}
         F(\l) = a \l^p + b \l
\label{bfit}
\end{equation}
to the $L=20$ data, where the linear term is motivated from the fact
that perturbative behavior (linear dependence on $\l$ ) is expected at
large $\l$ .  The fit gives the result shown in Fig.\ \ref{f_fit},
with an exponent $p=0.38$ which is not far off our guess of $p=0.42$.

\FIGURE[h!]{
\centerline{\includegraphics[width=8truecm]{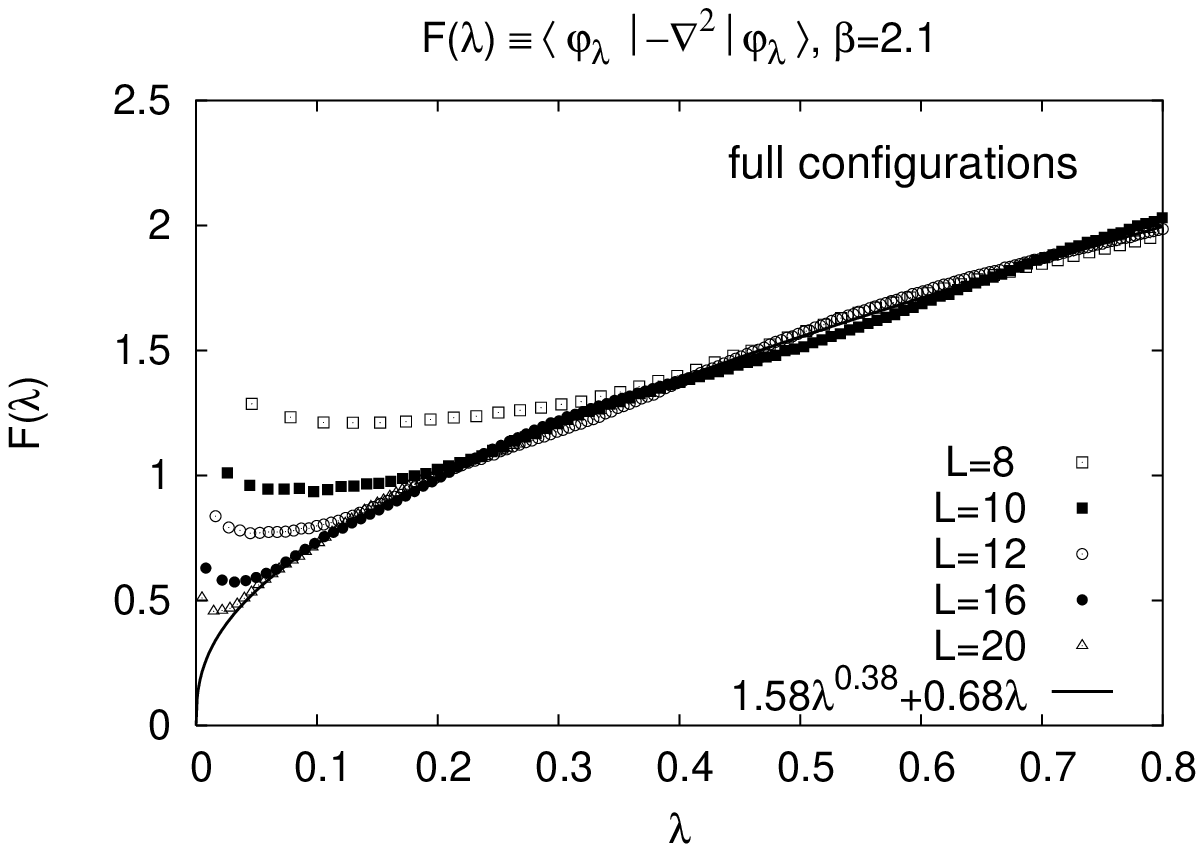}}
\caption{$F(\l)$ as in Fig.\ \protect\ref{f}, together with a best
fit to eq.\ (\protect\ref{bfit}).}\label{f_fit}}

\section{Proof of convexity of FMR in SU(2) lattice gauge theory}
\label{B}

The defining property \rf{definefmr} of the fundamental modular region
may be written
\begin{equation}
\sum_{xi} \mbox{Re Tr}\left[U_i(x) g_i(x)\right] \leq
\sum_{xi}\mbox{Re Tr}\; U_i(x),
\end{equation}
where $g_i(x) \equiv g(x+\hat{\imath})g(x)^{-1} $.  We write $g_i(x) =
c_i(x) + i\vec{\sigma}\cdot \vec{g}_i(x)$, where $c_i(x) = \pm [1 -
\vec{g}_i(x)^2]^{1/2}$, and the defining property reads,
\begin{equation}
\label{defprop}
- \sum_{xi} \vec{a}_i(x) \cdot \vec{g}_i(x) \leq \sum_{ix}
   b_i(x)\left[1 - c_i(x)\right],
\end{equation}
where the last factor is positive, $1 - c_i(x) \geq 0$.  Given that
this property holds for configurations $a_1$ and $a_2$,
\begin{eqnarray}
- \sum_{xi} \vec{a}_{1,i}(x) \cdot \vec{g}_i(x) &\leq& \sum_{xi}
b_{1,i}(x)\left[1 - c_i(x)\right],\\ - \sum_{xi} \vec{a}_{2,i}(x)
\cdot \vec{g}_i(x) &\leq& \sum_{xi} b_{2,i}(x)\left[1 - c_i(x)\right],
\end{eqnarray}
then we will show that it holds for $a\equiv\alpha a_1+\beta a_2$, as
stated in section V.  Upon multiplying the first equation by $\alpha$
and the second by $\beta$ and adding we obtain,
\begin{equation}
- \sum_{xi} \vec{a}_i(x) \cdot \vec{g}_i(x) \leq \sum_{xi}
  \left[\alpha b_{1,i}(x) + \beta b_{2,i}(x)\right]\left[1 -
  c_i(x)\right],
\end{equation}
where $\vec{a}_i(x) \equiv \alpha \vec{a}_{1,i}(x) + \beta
\vec{a}_{2,i}(x)$.  It follows from this that the defining property
\rf{defprop} that we wish to establish will be proven if we can show
that on each link $(xi)$ the inequality
\begin{equation}
  \alpha b_{1,i}(x) + \beta b_{2,i}(x) \leq b_i(x)
\end{equation}
holds, where $b_i(x) = [1 - \vec{a}_i(x)^2]^{1/2}.$

We introduce trigonometric functions
\begin{eqnarray}
b_{1,i}(x) = \cos \theta_{1,i}(x); &\ \ & \vec{a}_{1,i}(x) = \sin
\theta_{1,i}(x) \ \hat{n}_{1,i}(x),\\ b_{2,i}(x) = \cos
\theta_{2,i}(x); &\ \ & \vec{a}_{2,i}(x) = \sin \theta_{2,i}(x) \
\hat{n}_{2,i}(x),\\ b_i(x) = \cos \theta_i(x); &\ \ & \vec{a}_i(x) =
\sin \theta_i(x) \ \hat{n}_i(x),
\end{eqnarray}
where $0 \leq \theta_{1,i}(x) \leq \pi/2$, and $0 \leq \theta_{2,i}(x)
\leq \pi/2$, and $0 \leq \theta_i(x) \leq \pi/2$.  They are related by
\begin{equation}
\label{triangle}
  \alpha \sin \theta_i(x) \ \hat{n}_{1,i}(x) + \beta \sin
  \theta_{2,i}(x) \ \hat{n}_{2,i}(x) = \sin \theta_i(x) \hat{n}_i(x),
\end{equation}
and the property to be proved reads
\begin{equation}
\label{toshow}
  \alpha \cos \theta_{1,i}(x) + \beta \cos \theta_{2,i}(x) \leq \cos
  \theta_i(x).
\end{equation}
We have
\begin{equation}
   \alpha \cos \theta_{1,i}(x) + \beta \cos \theta_{2,i}(x) \leq
  \cos\left[\alpha \theta_{1,i}(x) + \beta \theta_{2,i}(x)\right],
\end{equation}
by the convexity of the cosine function, so it is sufficient to prove
\begin{equation}
\cos\left[\alpha \theta_{1,i}(x) + \beta \theta_{2,i}(x)\right] \leq
\cos \theta_i(x).
\end{equation}
The triangle inequality applied to \rf{triangle} reads
\begin{eqnarray}
\nonumber \sin \theta_i(x) &\leq& \alpha \sin \theta_{1,i}(x) + \beta
  \sin \theta_{2,i}(x)\\ &\leq& \sin \left[\alpha \theta_{1,i}(x) +
  \beta \theta_{2,i}(x)\right],
\end{eqnarray}
where we have also used the convexity of the sine function.  This
yields the inequality $\theta_i(x) \leq \alpha \theta_{1,i}(x) + \beta
\theta_{2,i}(x)$, from which \rf{toshow} follows.  This proves the
convexity of $\Lambda_+$.

\section{Renormalization-group flows toward Coulomb gauge}
\label{C}

         We compute the flow%
\footnote{The calculation reported here was motivated by
correspondence with Kurt Langfeld and Laurent Moyaerts.} of the gauge
parameter $a$ that appears in the interpolating gauge condition
\rf{interpolate} in the neighborhood of $a = 0$.

         Lorentz-invariance is not manifest in the interpolating gauge,
         and $A_0$ and ${\bf A}$ renormalize independently,
\begin{equation}
A_0 = Z_{A_0} A_0^R; \qquad {\bf A} = Z_{\bf A} {\bf A}^R,
\end{equation}
where superscript $R$ designates renormalized quantities.  The
renormalized gauge parameter $a^R$ is defined by
\begin{equation}
a = Z_a \ a^R,
\end{equation}
where the renormalization constant $Z_a$ is determined by the
renormalized gauge condition,
$$a^R \partial_0 A_0^R + \nabla \cdot {\bf A}^R = 0.$$ This gives
\begin{equation}
Z_a = Z_{A_0}^{-1} Z_{\bf A}.
\end{equation}

The $\gamma$-functions are defined by
\begin{eqnarray}
\nonumber \partial_t \ln Z_a &=& \gamma_a(g) , \\ \partial_t \ln
Z_{A_0} &=& \gamma_{A_0}(g), \\ \nonumber \partial_t \ln Z_{\bf A} &=&
\gamma_{\bf A}(g),
\end{eqnarray}
where $t \equiv \ln \Lambda$, and $\L$ is the ultraviolet cut-off.
The derivative is taken at fixed renormalized coupling constant $g^R$.
We have
\begin{equation}
\partial_t \ln Z_a = - \partial_t \ln Z_{A_0} + \partial_t \ln Z_{\bf
A},
\end{equation}
which gives
\begin{equation}
\gamma_a(g) = - \gamma_{A_0}(g) + \gamma_{\bf A}(g).
\end{equation}
From the perturbative expansions
\begin{eqnarray}
\nonumber Z_a &=& 1 + c_a \ln \Lambda \ (g^R)^2 + \dots,\\ Z_{A_0} &=&
1 + c_{A_0} \ln \Lambda \ (g^R)^2 + \dots, \\ \nonumber Z_{\bf A} &=&
1 + c_{\bf A} \ln \Lambda \ (g^R)^2 + \dots,
\end{eqnarray}
we obtain
\begin{eqnarray}
\nonumber \gamma_a &=& c_a \ g^2 + \dots, \\ \gamma_{A_0} &=& c_{A_0}
\ g^2 + \dots, \\ \nonumber \gamma_{\bf A} &=& c_{\bf A} \ g^2 +
\dots,
\end{eqnarray}
and
\begin{equation}
c_a = - c_{A_0} + c_{\bf A}.
\end{equation}

To one-loop order the renormalization constant satisfies
\begin{equation}
\partial_t \ln Z_a = c_a g^2 + \dots = { {c_a} \over {2 b_0 t} } +
\dots,
\end{equation}
where $b_0$ is the first coefficient of the $\beta$-function.  In pure
SU($N$) gauge theory it has the value
\begin{equation}
b_0 = { {11} \over {3} } { {N} \over {16\pi^2} }.
\end{equation}
We obtain, to one-loop order,
\begin{equation}
Z_a = {\rm const} \ t^{c_a/2b_0},
\end{equation}
which, with $t = \ln \L$, gives the leading dependence of the bare
gauge parameter on the cut-off $\Lambda$,
\begin{equation}
a(\L/\L_{\rm QCD}) = {\rm const} \ (\L/\L_{\rm QCD})^{c_a/2b_0}.
\end{equation}

Clearly $a = 0$ is a fixed point of the renormalization group.  To see
if it is also a stable fixed point, we may evaluate the coefficient
$c_a$ at $a = 0$, namely in Coulomb gauge, assuming that $c_a$ is
smooth in the neighborhood of $a = 0$.  The renormalization constants
$Z_{A_0}$ and $Z_{\bf A}$ in Coulomb gauge, are given to one-loop
order in eq. (B.37) of \cite{renormcoul}.  The coefficients have the
values
\begin{equation}
c_{A_0} = { {11} \over {6} } { {N} \over {8\pi^2} } \qquad c_{\bf A} =
{ {1} \over {2} } { {N} \over {8\pi^2} },
\end{equation}
which gives
\begin{equation}
c_a = - { {4} \over {3} } { {N} \over {8\pi^2} },
\end{equation}
and
\begin{equation}
a(\Lambda/\Lambda_{\rm QCD}) = { { {\rm const} } \over {
[\ln(\Lambda/\L_{\rm QCD})]^{4/11} } }.
\end{equation}

%
% References
%

\end{document}